\documentclass[journal,draftcls,onecolumn,12pt,twoside]{IEEEtranTCOM}

\usepackage{cite}

\ifCLASSINFOpdf
  \usepackage[pdftex]{graphicx}
\else

  \usepackage[dvips]{graphicx}
  \DeclareGraphicsExtensions{.eps}
\fi

\usepackage[cmex10]{amsmath}
\usepackage{amssymb}
\usepackage{algorithm}
\usepackage{algorithmic}
\usepackage{bbm}
\usepackage{mathrsfs}
\usepackage{dsfont}
\usepackage{array}
\usepackage{epstopdf}
\usepackage{setspace}
\usepackage{cite}
\usepackage{graphics}
\usepackage{graphicx}
\usepackage{multirow}
\usepackage{xcolor}
\usepackage{stfloats}
\usepackage{bm}
\usepackage{subfigure}
\usepackage{pdflscape}
\IEEEoverridecommandlockouts

\begin{document}

\title{Caching with Time Domain Buffer Sharing}

\author{Wei~Chen,~\IEEEmembership{Senior Member,~IEEE},~and~H. Vincent Poor,~\IEEEmembership{Fellow,~IEEE}

\thanks{Wei Chen is with the Department of Electronic Engineering and Beijing National Research Center for Information Science and Technology, Tsinghua University, Beijing, 100084 China, e-mail: wchen@tsinghua.edu.cn.}

\thanks{H. Vincent Poor is with the Department of Electrical Engineering, Princeton University, New Jersey, 08544, USA, e-mail: poor@Princeton.EDU.}

\thanks{This research was supported in part by the U.S. National Science Foundation under Grants CNS-1702808 and ECCS-1647198, and the National Natural Science Foundation of China under Project Nos. 61671269 and 61621091.}
}

\markboth{Submitted paper}%
{Submitted paper}

\maketitle

\setstretch{1.6}

\begin{abstract}
In this paper, storage efficient caching based on time domain buffer sharing is considered. The caching policy allows a user to determine whether and how long it should cache a content item according to the prediction of its random request time, also referred to as the request delay information (RDI). In particular, the aim is to maximize the caching gain for communications while limiting its storage cost. To achieve this goal, a queueing theoretic model for caching with infinite buffers is first formulated, in which Little's law is adopted to obtain the tradeoff between the hit ratio and the average buffer consumption. When there exist multiple content classes with different RDIs, the storage efficiency is further optimized by carefully allocating the storage cost. For more practical finite-buffer caching, a $G/GI/L/0$ queue model is formulated, in which a diffusion approximation and Erlang-B formula are adopted to determine the buffer overflow probability and the corresponding hit ratio. The optimal hit ratio is shown to be limited by the demand probability and buffer size for large and small buffers respectively. In practice, a user may exploit probabilistic caching with random maximum caching time and arithmetic caching without any need for content arrival statistics to efficiently harvest content files in air.
\end{abstract}


\IEEEpeerreviewmaketitle

\newpage

\setstretch{1.6}

\section{Introduction}

The explosive growth of mobile multimedia and social networking applications has stimulated a corresponding explosively increasing demand for bandwidth in the emerging fifth-generation of mobile systems (5G). However, standard on-demand transmission infrastructures can hardly cope with the dramatic increase of mobile data traffic, due to the scarcity of wireless resources such as radio spectrum, constrained average or peak power, and limited base station sites. Moreover, the efficiencies of many radio resources have already achieved their theoretical limits after being heavily exploited in the past decades.

Caching holds the promise of trading cheap storage resources for substantial throughput increases in content-centric networks. By allowing a user to cache popular content items before requested, the peak data rate can be significantly reduced. Therefore, caching becomes a key solution in the era of 5G to meet the stringent requirements of data rate, energy efficiency, and Quality-of-Service (QoS) \cite{Leung} - \cite{Sherman}. To fully exploit the caching gain, context-awareness is enabled in paradigm-shift 5G network architectures, where social networking \cite{Social}, big data analytic \cite{Debbah_Bigdata}, recommendation systems \cite{Recommend}, and natural language processing \cite{W.Chen_WCM} are adopted for popularity prediction. Then the synergy between communication, caching, and computing is regarded as the cornerstone in building versatile 5G-grade deployments \cite{3C_Synergy}, \cite{3C_Cornerstone}.

The applications of caching in 5G motivate an extensive study on the communication-storage tradeoff that characterizes a fundamental limit of caching. Maddah-Ali and Niesen first revealed the tradeoffs between the delivery rate and memory size for uniform demands \cite{M.Ali} and nonuniform demands \cite{Ali_Nonuniform}. The rate-memory tradeoff of coded caching for multilevel popularity was presented in \cite{Diggavi}. Aiming at characterizing the memory-rate tradeoff better, Sengupta and Tandon proposed a tighter lower bound on the worst-case delivery rate in \cite{Approx_Tradeoff}. To further improve the rate-memory tradeoff, Amiri and G\"und\"uz conceived an enhanced coded caching, in which content items are partitioned into smaller chunks \cite{Gunduz_Improve}. When the storage resources are shared, how to efficiently allocate them among multiple users becomes a critical issue. Optimal storage allocations were proposed for wireless cloud caching systems by Hong and Choi \cite{W.Choi}, and for heterogeneous caching networks by Vu, Chatzinotas, and Ottersten \cite{B.Ottersten}. In wireless device-to-device networks, caching is also expected to bring dramatic capacity gain. Given cache size constraint, the capacity upper and lower bounds were revealed by Ji, Caire, and Molisch in \cite{Caire}. More recently, scaling laws for popularity-aware caching with limited buffer size were found by Qiu and Cao in \cite{G.Cao}.

The demand probability is shown to play a key role in the rate-memory tradeoff. In most existing works, it is assumed to be time-invariant. In practice, however, content popularity may not remain unchanged. More particularly, a content item may become more popular, e.g. due to information propagation in social networks, but will finally be expired at the end of its lifetime. As a result, time-varying content popularity models were proposed in \cite{H.Liu}, \cite{A.Eryilmaz}, and \cite{W.Chen_TCOM1}. In our previous work \cite{W.Chen_TCOM1}, we were aware of a fact that most data is requested only once. Based upon this observation, we defined the request delay information (RDI) as the probability density function ($p.d.f.$) of a user's request time/delay for a content item. In practice, the RDI can be estimated from a content item's labels or key words and then applied to to predict the user's request time for a content item. The communication-storage tradeoffs of caching with various RDI were revealed in \cite{W.Chen_TCOM1} and \cite{W.Chen_TCOM2} for unicast and multicast transmissions respectively.

The time-varying popularity allows a user to remove a cached content item from its buffer when this content item becomes outdated or less popular.\footnote{In this paper, the two terms buffer and memory are used interchangeably.} Once a content item is removed, the occupied buffer space can be released in order to cache other content items. As such, the user may reuse its buffer in the time domain. When time domain buffer sharing is enabled, recent works \cite{W.Chen_Globecom} and \cite{Retention} show that the storage cost is also an increasing function of the content caching time. In general, however, how to efficiently share the buffer in the time domain still remains as an open but fundamental problem.

In this paper, we are interested in storage efficient caching based upon time domain buffer sharing. Our aim is to maximize the caching gain for communications, while limiting the storage cost. In particular, we shall determine whether and how long a user should cache a content item, according to its RDI. We first focus on the infinite-buffer assumption that allows us to formulate an infinite-buffer queue model. In the queueing-theoretic model, Little's law \cite{Little} is adopted to bridge the storage cost and the maximum caching time, thereby giving the communication-storage tradeoff. To strike the optimal tradeoff between the hit ratio and the average buffer consumption, we present two storage efficient caching policies for the homogenous and heterogenous RDI cases, in which all content items have the same and different RDI respectively. With homogenous RDI, we conceive a probabilistic caching policy in which the random maximum caching time obeys a certain distribution. With heterogenous RDI, we allocate the storage cost among different content classes to maximize the overall storage efficiency. The storage allocation is formulated as a convex optimization, the optimal solution of which has a simple structural result. The solution also leads to a decentralized arithmetic caching without any need for the global knowledge of content arrival process. It allows a user to efficiently harvest popular content items in air and decide how long a content item should be cached, simply based on the user's local RDI prediction.

The communication-storage tradeoff under the infinite-buffer assumption lays a theoretical foundation for storage efficient caching with finite buffer, which is more practical in general. In this scenario, the hit ratio is jointly determined by that of the infinite-buffer caching, as well as, the blocking probability due to buffer overflow. To obtain the blocking probability, we formulate a $G/GI/L/0$ queue model and then apply the diffusion approximation. By this means, we show that the hit ratio maximization is approximately equivalent to a one-dimensional quasi-convex optimization, the variable of which is the mean caching time. Two approximate but analytical solutions are presented for large and small buffers, in which the hit ratios are limited by the demand probability and buffer size respectively. For decentralized implementation without statistics of content arrivals, arithmetic caching with finite buffer is also conceived.

The rest of this paper is organized as follows. Section II presents the system model. Based upon the infinite buffer assumption, Sections III and IV investigate the storage efficient caching with homogeneous and heterogeneous RDIs respectively. In Section V, the more practical finite-buffer caching is presented. Finally, numerical results and conclusions are given in Sections VI and VII, respectively.

\section{System Model}

\begin{figure}[!t]
\centering
\includegraphics[width=6in]{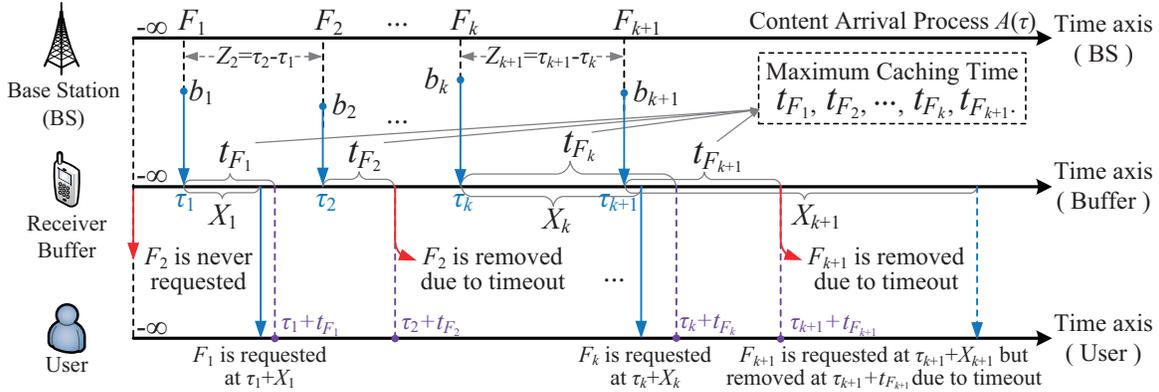}
\caption{Caching with time domain buffer sharing.}
\label{fig_sys}
\end{figure}

Consider a caching system, as shown in Fig. \ref{fig_sys}. The $k$th ($k \in \mathbbm N ^+$) content item transmitted by the base station (BS) is denoted by $F_k$, which consists of $b_k$ bits.\footnote{Throughout this paper, the terms content file and content item are used interchangeably.} The transmission of $F_k$ is accomplished at time $\tau_k$. From the receiver's perspective, the content arrival process is defined as $A(\tau) = \sum_{k=1}^{\infty} \mathbbm 1 \{ \tau_k \leq \tau \}$, with content interarrival time given by $Z_k = \tau_k - \tau_{k-1}$. Assuming that the limit $\underset{K \to \infty} {\lim} \frac{1}{K} \sum_{k=1}^{K} Z_k $ exists, we present the content arrival rate as $\lambda = \left( \underset{K \to \infty} {\lim} \frac{1}{K} \sum_{k=1}^{K} Z_k \right)^{-1} $.

The user may ask for content file $F_k$ after a random delay $X_k$ from $F_k$'s being sent. If $F_k$ will never be requested by the user, then we set $X_k = - \infty$. The $p.d.f.$ of $X_k$ is denoted by $p_{X_k}(x)$, also referred to as the RDI of $F_k$. We categorize content items into $N$ classes according to their statistical RDI. Let $\mathcal F_i$, $i = 1, \ldots, N$, denote the $i$th class, where all the content items $F_k \in \mathcal F_i$ have the same statistical RDI denoted by $p_i(x)$, i.e., $p_{X_k}(x) = p_i(x)$. If $N=1$, the content flow has homogenous RDI. In this case, the class index $i$ is dropped. Otherwise, the content flow has heterogenous RDI. Moreover, the cumulative distribution function ($c.d.f.$) of $p_i(x)$ is denoted by $P_i(x) = \int_{-\infty}^{x}p_i(z)dz$. In this context, the probability that a content item of class $\mathcal F_i$ will not be requested after transmission is given by $q_i = P_i(0)$. In other words, its demand probability is $(1-q_i)$. Furthermore, the minimum and maximum possible request delay are presented by $t_i^{\inf} = \sup \{x | p(x) = 0, x \geq 0 \}$ and $t_i^{\sup} = \sup \{x | p(x) > 0, x \geq 0 \}$, respectively. Let $\nu_i$ stand for the conditional expectation of $X_i$ given $X_i \geq 0$, i.e., $\nu_i = \mathbbm E \{X_i |X_i \geq 0\}$ or $\nu_i = \int_{0}^{\infty} x p_i(x)dx$.

We assume that the limit $B_i = \underset{K \to \infty} {\lim} \frac{\sum_{k=1}^{K} b_k \mathbbm 1_{\{F_k \in \mathcal F_i\} }}{\sum_{k=1}^{K} \mathbbm 1_{\{F_k \in \mathcal F_i\}} }$ exists for all $i = 1, \ldots, N$, which represent the average number of bits per content item of class $\mathcal F_i$. Let $\hat \pi_i$ denote the probability that a content item belongs to class $\mathcal F_i$. Therefore, the average number of bits per content item is obtained by $B = \sum_{i=1}^{K} \hat \pi_i B_i$ (bits). The overall transmission rate of the base station is given by $\lambda B$ (bits/second). Furthermore, the probability that a bit belongs to class $\mathcal F_i$ is determined by $\pi_i = \frac{\hat \pi_i B_i}{B}$.

Next, we present the performance metrics of the communication gain and storage cost of caching. In particular, the caching gain for communications is characterized by the effective throughput, which is defined to be the average number of bits that the user reads from its buffer in unit time. Since the effective throughput increases linearly with the hit ratio, the hit ratio is regarded as an alternative performance metric for the caching gain. Let $R_i$ denote the effective throughput contributed by content items of flow $\mathcal F_i$. Therefore, the sum effective throughput or the overall caching reward is obtained by $R = \sum_{i=1}^{N} R_i$. The storage cost is characterized by two different performance metrics that are applicable in the infinite and finite buffer scenarios respectively. When the user is equipped with an infinite buffer, the storage cost is the average buffer consumption that is defined as the average number of bits cached in the receiver buffer. Let $S_i$ denote the average number of cached bits from flow $\mathcal F_i$. Then the overall storage cost is obtained by $S = \sum_{i=1}^{N} S_i$. When the user is equipped with a finite buffer, the storage cost is simply the buffer size.

Finally, our aim is to efficiently share or reuse the receiver buffer in the time domain. To this end, we shall carefully decide whether and how long the user should cache a content item. Intuitively, a content file should be removed from the buffer if it has been requested and read, because in practice a BS is seldom asked to transmit the same content file to a user twice.\footnote{If the user thinks he or she will read a content file again, this content file can be stored locally. However, such storage cost is beyond the scope of this paper.} On the other hand, we should carefully limit the caching time of each content item in order for efficiently sharing the buffer in the time domain. To achieve this, let $t_{F_k}$ denote the maximum caching time for content file $F_k$. In other words, $F_k$ must be removed from the user's buffer after being cached for time $t_{F_k}$, even though it has not been read.

A simple caching policy is static caching, in which all content items of flow $\mathcal F_i$ have the same maximum caching time, i.e., $t_{F_k} = t_i$ for all $F_i \in \mathcal F_i$. A generalized caching policy is probabilistic caching, in which $t_i$ can be a non-negative random variable. Let $f_i(t_i)$ denote the $p.d.f.$ of the maximum caching time $t_i$ of content class $\mathcal F_i$. When $f_i(x) = \delta(x-t_i)$, the probabilistic caching reduces to the static caching with deterministic maximum caching time $t_i$. In other words, static caching is a special case of probabilistic caching. In this work, our main target is to find the optimal maximum caching time $t_i$ or its probability density function $f_i(t_i)$ to maximize the effective throughput under the constraint of the storage cost.

\section{Caching with Infinite Buffer and Homogenous RDI}
In this section, we focus on caching with infinite buffer and homogenous RDI, thereby dropping the content class index $i$. Based on the effective throughput and storage cost analysis of static caching policies, we present a normalized rate-cost function, which characterizes the communication-storage tradeoff. A probabilistic caching policy is further conceived to achieve the optimal rate-cost tradeoff.

\subsection{Effective Throughput and Storage Cost of Static Caching}
In this subsection, we consider static caching policy with a fixed maximum caching time $t$. The effective throughput and the storage cost are presented as two increasing functions of $t$.

\newtheorem{lemma}{Lemma}
\begin{lemma}
\label{lem_R}
The effective throughput is given by
\begin{equation}
\label{R}
R(t) = \lambda B \int_{0}^{t} p(x)dx.
\end{equation}
\end{lemma}
\begin{IEEEproof}
Given the $p.d.f$ of the request delay, $p(x)$, and the maximum caching time $t$, the hit ratio is obtained by
\begin{equation}
\label{hit_rate}
r(t) = \int_{0}^{t} p(x)dx.
\end{equation}
Since the effective throughput is equal to $\lambda B r(t)$, the lemma follows.
\end{IEEEproof}

\begin{figure}[!t]
\centering
\includegraphics[width=6in]{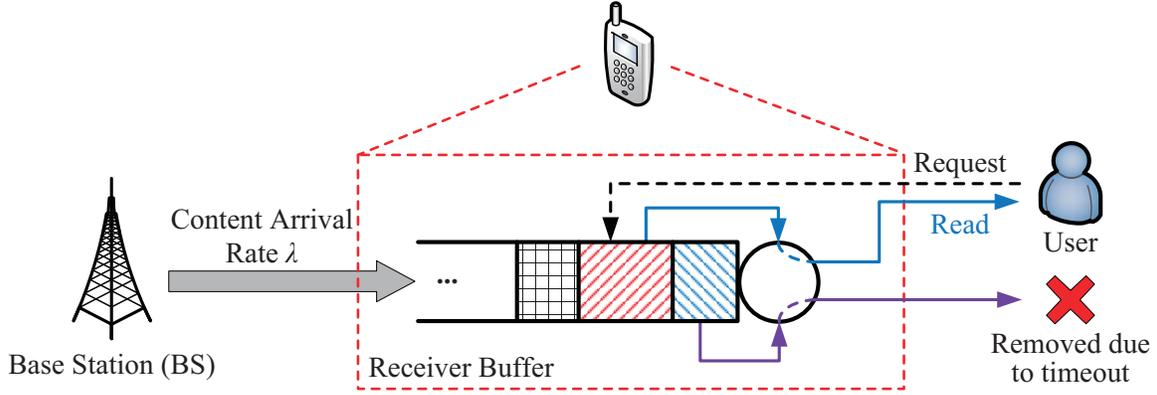}
\caption{A queueing-theoretic model of infinite buffer.}
\label{fig_queue_infinite}
\end{figure}

Next, we analyze the storage cost by formulating  a queueing-theoretic model as shown in Fig. \ref{fig_queue_infinite}, in which a content file arrives after being transmitted and departs when it is removed. Although it is clearly not a first-in-first-out (FIFO) queue, Little's law \cite{Little} is still applicable, which gives the following lemma.

\begin{lemma}
\label{lem_S}
The average buffer consumption is given by
\begin{equation}
\label{S}
S(t) = \lambda B \left[ \int_{0}^{t} xp(x) dx + t \left( q + \int_{t}^{\infty} p(x)dx \right) \right].
\end{equation}
\end{lemma}
\begin{IEEEproof}
The average number of content files cached in the buffer is equal to the average queue length $Q$. According to Little's law, we have $Q = \lambda \mathbbm E \{ W \} $, in which $W$ denotes the random caching time of a content file. Consider a content file with request delay $X$. We have $W = X$ if this file is requested before its maximum caching time, i.e., $0 \leq X < t$. Otherwise, we have $W=t$. Therefore, its caching time is given by
\begin{equation}
\label{waiting}
W = X \mathbbm 1_{\{ 0 \leq X < t \}} + t \left(1-\mathbbm 1_{\{ 0 \leq X < t \}} \right),
\end{equation}
the expectation of which is then determined by
\begin{equation}
\label{mean_W}
\mathbbm E \{ W \} = \int_{0}^{t}xp(x)dx + t \left(\int_{-\infty}^{0} + \int_{t}^{\infty} \right) p(x)dx.
\end{equation}
Recalling that the probability that a content item will never be requested is $q=P(0)$, we get Eq. (\ref{S}).
\end{IEEEproof}

Lemmas \ref{lem_R} and \ref{lem_S} imply that both the effective throughput $R(t)$ and storage cost $S(t)$ monotonically increase with the maximum caching time $t$. As a result, there exists a fundamental tradeoff between them, as shown in the following theorem.
\newtheorem{theorem}{Theorem}
\begin{theorem}
\label{thm_tradeoff}
The storage cost is an increasing function of the effective throughput given by
\begin{equation}
\label{tradeoff}
S(R) =
\lambda B \left[ \int_{0}^{P^{-1}\left( \frac{R}{\lambda B} + q \right)} x d P(x) + \left( 1 - \frac{R}{\lambda B} \right) P^{-1}\left( \frac{R}{\lambda B} + q \right) \right],
\end{equation}
where $P^{-1}(z) = \inf\{x | P(x) \geq z\}$, which is the inverse function of $P(x)$ when $P(x)$ is continuous.
\end{theorem}
\begin{IEEEproof}
Since Eq. (\ref{R}) can be rewritten as $R = \lambda B \left( P(t) - q \right)$, the maximum caching time is given by
\begin{equation}
\label{t_R}
t = P^{-1}\left( \frac{R}{\lambda B} + q \right).
\end{equation}
By substituting Eq. (\ref{t_R}) and $\int_{t}^{\infty} p(x)dx = 1 - P(t)$ into Eq. (\ref{S}), we obtain Eq. (\ref{tradeoff}).
\end{IEEEproof}

\subsection{Normalized Rate-Cost Function of Static Caching}
In this subsection, we normalize both the storage cost $S$ and the effective throughput $R$ by the BS's throughput given by $\lambda B$. More specifically, the normalized storage cost is defined as $s = \frac{S}{\lambda B}$. From Eqs. (\ref{S}) and (\ref{mean_W}), we see that the normalized storage cost is equal to the mean caching time, i.e., $s = \mathbbm E \{ W \}$. From Eqs. (\ref{R}) and (\ref{hit_rate}), we also see that the normalized effective throughput $\frac{R}{\lambda B}$ is equal to the hit ratio defined by Eq. (\ref{hit_rate}), i.e., $r = \frac{R}{\lambda B}$. From Eq. (\ref{tradeoff}), we may write the normalized rate-cost function as
\begin{equation}
\label{norm_sr}
s(r) = \int_{0}^{P^{-1}( r + q )} x d P(x) + ( 1 - r ) P^{-1}( r + q ),
\end{equation}
the domain of which is $r \in (0, 1-q] $. From Eq. (\ref{S}), one can see that the codomain of $s(r)$ is $s(r) \in (t^{\inf}, \nu + q t^{\sup}]$ if $\nu < \infty$ and $qt^{\sup} < \infty$. Otherwise, we have $s(r) \in (t^{\inf}, \infty)$. Since $\frac{R}{S(R)} = \frac{r}{s(r)}$, the storage efficiency is irrelevant to the BS's transmission rate $\lambda B$ but simply relies on the statistical RDI $p(x)$. As a result, the tradeoff between the effective throughput and the storage cost is equivalent to the tradeoff between the hit ratio and the mean caching time.

Since $s(r)$ relies only on $p(x)$, it can be regarded as a transform from $p(x)$, which we shall refer to as the cost-$p.d.f.$ transform (CP transform for brevity), denoted by $p(x) \overset{\mathscr C}{\leftrightarrow} s(r)$ or $s(r) = \mathscr {C}[p(x)]$. It is easy to check that $\mathscr {C}[p(x)] = \mathscr {C}[p(x) \mathbbm 1_{\{ x \geq 0 \}} + q \delta(-\infty) \mathbbm 1_{\{ x < 0 \}}]$ for any $p(x)$. Hence, any $p(x)$ is equivalent to its normalized form given by $p(x) \mathbbm 1_{\{ x \geq 0 \}} + q \delta(-\infty) \mathbbm 1_{\{ x < 0 \}}$, on which we shall focus in the remainder of this paper. Tables \ref{Typical_pdf} and \ref{CP_property} present the CP-transforms of five typical $p.d.f.$s, as well as, some interesting properties of the CP-transform to calculate or bound $s(r)$ easily based on the CP-transform of a standard $p(x)$. From Table \ref{Typical_pdf}, we see $s(r) = \frac{r}{\iota}$ when $p(x) = \iota e^{-\iota x}$. This is not surprising because the exponential distribution is memoryless. In the following, we show it is not only sufficient but also necessary.

\newtheorem{corollary}{Corollary}
\begin{corollary}
\label{cor_linear_exp}
$s(r) = \frac{r}{\iota}$ if and only if $p(x) = \iota e^{-\iota x}$.
\end{corollary}
\begin{IEEEproof}
Our proof relies on the observation that $r'(t) = P'(t)$ and $s'(t) = 1+q - P(t)$. Therefore, $s(r) = \frac{r}{\iota}$ if and only if $\iota^{-1} P'(t) + P(t) - (1 + q ) = 0$. The solution to this first-order linear ordinary differential equation is $P(t) = (1+q)\left(1 - e^{-\iota (t - \hat \iota)} \right)$. Recalling the boundary value conditions $P(0)=q$ and $P(\infty) = 1$, we have $q=0$ and $\hat \iota=0$, which completes our proof.
\end{IEEEproof}

\begin{table}[!t]
	\small
	
	\setstretch{1}
	
	\renewcommand{\arraystretch}{1.6}
	
	

	\caption{CP-transforms of typical $p.d.f.$s}
	
	\label{Typical_pdf}
	
	\centering
	
	\begin{tabular}{p{1.7cm}|p{6.6cm}|p{6cm}}
		\hline
		\text{Distributions}  &  \textit{p.d.f}  &  \text{CP transforms}\\
		
		\hline
		\text{Exponential}  &		
		$\iota e^{-\iota x}$,  $x\in[0,+\infty)$  &
		$\frac{r}{\iota}$\\
		
		\hline
		\text{Uniform}  &		
		$\frac{1}{\iota_2-\iota_1}$,  $x\in[\iota_1,\iota_2]$  &		
		$-\frac{1}{2}(\iota_2-\iota_1)r^2 + (\iota_2-\iota_1)r+\iota_1$\\
		
		\hline
		\text{Triangular}  &
		$\frac{2 (x-\iota_1) }{\hat \iota} \mathbbm{1}_{\{ x \le \iota_3 \}} + \frac{2 (\iota_2-x) }{\tilde \iota} \mathbbm{1}_{\{ x > \iota_3 \}}$,  $x\in[\iota_1,\iota_2]$
		 \newline
		$\hat \iota = (\iota_2-\iota_1)(\iota_3-\iota_1)$,
		$\tilde \iota = (\iota_2-\iota_1)(\iota_2-\iota_3)$  &
		$\begin{cases}
			\sqrt{\hat \iota}\left( r^{\frac{1}{2}} -  \frac{1}{3}r^{\frac{3}{2}}\right) +\iota_1 & r \le \frac{\iota_3-\iota_1}{\iota_2-\iota_1}\\
			\frac{1}{3}\sum_{i=1}^3\iota_i - \frac{1}{3}\sqrt{\tilde \iota}(1-r)^{\frac{3}{2}}& r > \frac{\iota_3-\iota_1}{\iota_2-\iota_1}
		\end{cases}$\\
		
		\hline
		\text{Pareto}  &
		$\frac{\iota_1 \iota_2^{\iota_1}}{x^{\iota_1+1}}$,  $x\in[ \iota_2, +\infty )$  &
		$\begin{cases}
			\frac{\iota_2}{1-\iota_1}(1-r)^{\frac{\iota_1-1}{\iota_1}}- \frac{\iota_2}{1-\iota_1}+\iota_2  & \iota_1 \neq 1\\
			\iota_2 \ln\frac{1}{1-r} +\iota_2 &\iota_1 = 1
		\end{cases}$\\
		
		\hline
		\text{Arcsine}  &
		$\frac{1}{\pi} \sqrt{\frac{1}{x(\iota - x)}}$, $x\in (0,\iota)$  &
		$\iota \left[ \left(1-r\right) \sin^2 \left(\frac{\pi}{2} r\right) + \frac{1}{2} r-\frac{1}{2\pi} \sin(\pi r)\right]$\\
		
		\hline
		
	\end{tabular}
	
\end{table}

\begin{table}[!t]
\small
	\setstretch{1}
	\renewcommand{\arraystretch}{1.5}
	\caption{Key Properties of the CP Transform}
	\label{CP_property}
	\newcommand{\tabincell}[2]{\begin{tabular}{@{}#1@{}}#2\end{tabular}}
	\begin{tabular}{p{3.3cm}|p{4.5cm}|p{7cm}}
		\hline
		\text{Operators}  &  $p.d.f.$  &  \text{CP transforms}\\
		
		\hline
		
		\text{Time scaling,} $\Psi_\xi^{1}$  &  $p(\xi x)$  &
		$\xi^{-1} s(r)$\\
		
		\text{Time shift,} $\Psi_\xi^{2}$ &  $p(x - \xi)$  &
		$ s(r) + \xi$\\
		
		\text{Density scaling,} $\Psi_\xi^{3}$  &
		$\xi p(x) + (1 - \xi)\delta(-\infty)$  &
		$\xi s\left(\frac{r}{\xi}\right) + (1-\xi)P^{-1}\left( \frac{r}{\xi} + q \right)$\\
		
		\text{Rate shift,} $\Psi_\xi^{4}$   &
		$ p(x) + \xi\delta(x- \zeta) + (q-\xi) \delta(-\infty)$ &
		$s(r) \mathbbm 1_{\{ x < \zeta \}} + \big\{ s(r-\xi) + \xi \big[ \zeta - P^{-1}(r+q-\xi) \big] \big\} \mathbbm 1_{\{ x \geq \zeta \}}$\\
		
		\hline
		
		\text{Stochastic order relation}  &
		$P_1(x) \leq P_2(x)$, $\forall x \geq 0$  &
		$r_1(s) \geq r_2(s)$\\
		
		\text{Linear combination}  &
		$p(x) = \theta p_1(x) + (1-\theta) p_2(x)$  &
		$\min \left\{ r_1(s), r_2(s) \right\} \leq r(s) \leq \max \left\{ r_1(s), r_2(s) \right\}$\\
		
		\hline
	\end{tabular}
\end{table}

\subsection{Probabilistic Caching with Homogeneous RDI}
In this subsection, we shall show that the rate-cost tradeoff can be further optimized by the probabilistic caching policies, due to the following observation.
\begin{lemma}
\label{lem_convex}
If two rate-cost pairs $\left(s(r), r\right)$ and $\left(s(\tilde r), \tilde r\right)$ are achievable, then the rate-cost pair given by $\left( s(\theta r + (1-\theta) \tilde r), \theta r + (1 - \theta) \tilde r \right)$ is also achievable for $0 \leq \theta \leq 1$.
\end{lemma}
\begin{IEEEproof}
The desired rate-cost pair is achieved by randomly choosing the maximum caching time $t$ for each content file, i.e., $t = P^{-1}(r+q)$ with probability $\theta$ and $\tilde t = P^{-1}(\tilde r+q)$ with probability $(1-\theta)$.
\end{IEEEproof}

By applying probabilistic caching, any linear combination of the achievable rate-cost pairs $\left(s(r), r\right)$ in the rate-cost curve characterized by Eq. (\ref{norm_sr}) is also achievable. Furthermore, since there is no need to cache any content item when $r = 0$, the storage cost is $0$. In summary, the following theorem holds.
\begin{theorem}
\label{thm_envelop}
The optimal rate-cost function is the lower convex envelope of $s(r)$ and $s(0) = 0$ given by
\begin{equation}
\label{envelope}
\begin{split}
& \breve s(r) =  \\
& \sup \Bigg \{s_e(r) \Bigg | 
s_e''(r) \geq 0, s_e(0) = 0, s_e(r) \leq
\int_{0}^{P^{-1}( r + q )} x d P(x) + ( 1 - r ) P^{-1}( r + q ), r \in (0, 1-q]
\Bigg \}.
\end{split}
\nonumber
\end{equation}
\end{theorem}

According to Theorem \ref{thm_envelop}, the probabilistic caching policy brings a storage efficiency gain if and only if $\breve s(r) \neq s(r)$, or more specifically, $s(r)$ is non-convex or $\underset {r \to 0} {\lim} s(r) \neq 0$. To shed some new light on $\breve s(r)$, we are interested in the concavity or convexity of $s(r)$. Consider $p(x)$ that is differentiable for $x \in [t^{\inf}, t^{\sup}]$. Then Fa\`{a} di Bruno's formula gives\footnote{Eq. (\ref{derivative2}) is also obtained by applying \emph{Example 29} in Chapter 5 of \cite{Zorich}.}
\begin{equation}
\label{derivative2}
s''(r) = - \frac {p^2(t) + (1-r)p'(t)}{p^3(t)},
\end{equation}
where $t=P^{-1}(r+q)$. Eq. (\ref{derivative2}) implies that the concavity and convexity of $s(t)$ is determined by $p^2(t) + (1-r)p'(t)$. Once $s(r)$ becomes either concave or convex, the optimal rate-cost function can be significantly simplified, as shown in the following theorem.

\begin{theorem}
\label{thm_cvx_criterion}
The optimal rate-cost function is given by
\begin{equation}
\label{simplify}
\breve s(r) = \\
\left\{
\begin{array}{ll}
\alpha r      &\textrm{if } p'(t) \geq \frac{p^2(t)}{r(t)-1}, \forall t \in [t^{\inf}, t^{\sup}],  \\
s(r)          &\textrm{if } p'(t) \leq \frac{p^2(t)}{r(t)-1}, \forall t \in [0, t^{\sup}] , t^{\inf} = 0,
\end{array}
\right.
\end{equation}
where $\alpha = \frac{\nu + q t^{\sup}}{1-q}$.
\end{theorem}

\begin{IEEEproof}
When $s''(r) \leq 0$, $s(r)$ is a concave function satisfying $s(1) = \nu$. Since $s(0) = 0$, the lower convex envelope of $s(t)$ is a straight line joining the points $(0,0)$ and $(1-q,\mu+q t^{\sup})$, the slope of which is $\alpha$. When $s''(r) \geq 0$, $s(r)$ is concave. Moreover, we have $\underset {r \to 0} {\lim} s(r) = 0$ because $t^{\inf} = 0$. As a result, the lower convex envelope of $s(r)$ is itself.
\end{IEEEproof}

Recalling that $q \leq 1-r \leq 1$, we may obtain a sufficient but not necessary condition for the concavity or convexity of $s(t)$, which further simplifies Eq. (\ref{simplify}) to be
\begin{equation}
\label{more_simplify}
\breve s(r)=\left\{
\begin{array}{ll}
\alpha r             &\textrm{if } p'(t) \geq - p^2(t), \forall t \in [t^{\inf}, t^{\sup}], \\
s(r)                 &\textrm{if } p'(t) \leq - \frac{p^2(t)}{q}, \forall t \in [0, t^{\sup}],  t^{\inf} = 0.
\end{array}
\right.
\end{equation}
Clearly, $\breve s(r) = \alpha r$ if $p(x)$ is a nondecreasing function on the interval $[t^{\inf}, t^{\sup}]$.

\section{Caching with Infinite Buffer and Heterogenous RDI}
In this section, we focus on the infinite buffer scenario with heterogenous RDI, where there exist multiple content classes $\mathcal F_i$, $i = 1, \ldots, N$. To distinguish them, a flow index $i$ is assigned to the notations developed in the previous section. In this case, how to efficiently allocate storage resources among various content classes becomes a critical issue.

\subsection{Joint Rate-Cost Allocation}
Our aim is minimize the overall storage cost given a target normalized effective throughput, denoted by $\mathsf r$, which is feasible when $0 \leq \mathsf r < \sum_{i=1}^{N} \pi_i (1-q_i)$. Since $\pi_i$ denotes the probability that a cached bit belongs to class $\mathcal F_i$, the normalized effective throughput and storage cost are expressed as $\sum_{i=1}^{N} \pi_i r_i$ and $\sum_{i=1}^{N} \pi_i \breve s_i(r_i)$, respectively. Moreover, we have the effective throughput $\lambda B \sum_{i=1}^{N} \pi_i r_i$ and the average buffer consumption $\lambda B \sum_{i=1}^{N} \pi_i \breve s_i(r_i)$. Therefore, the joint rate-cost allocation is equivalent to the joint normalized rate-cost allocation formulated as
\begin{equation}
\label{cvx}
\begin{array}{rrcl}
\displaystyle \min & \multicolumn{3}{l}{ \sum_{i=1}^{N} \pi_i \breve s_i(r_i) } \\
\textrm{s.t.} & \sum_{i=1}^{N} \pi_i r_i & =    & \mathsf r   \\
              &                      r_i & \in  & [0,1-q_i].
\end{array}
\end{equation}

Problem (\ref{cvx}) is a convex optimization problem that can be solved in low complexity, because the rate-cost functions $\breve s_i(r_i)$ are convex for $i = 1, \ldots, N$. Let $r_i = \varsigma_i(\beta)$ denote the inverse function of the derivative of the normalized rate-cost function of class $\mathcal F_i$, i.e., $\breve s_i'(\varsigma_i(\beta)) = \beta$.\footnote{If $\breve s(r) = s(r)$, we have $\breve s'(r) = \frac{1-r}{p\left(P^{-1}(r+q)\right)}$. If $\breve s(r) = \alpha r$, then $\breve s'(r) = \alpha$.} Then a structural result for problem (\ref{cvx}) is presented in the following theorem.

\begin{theorem}
\label{solution}
The optimal solution to (\ref{cvx}) is given by
\begin{equation}
\label{r_opt}
r_i^*(\beta) = \left\{
\begin{array}{cl}
1 - q_i             &\textrm{ if    } \breve s_i'(1-q_i) \leq \beta \\
\varsigma_i (\beta) &\textrm{ otherwise},
\end{array}
\right.
\end{equation}
where $\beta$ is a positive number satisfying $\sum_{i=1}^{N} \pi_i r_i^*(\beta) = \mathsf r$.\footnote{A careful reader may notice that $\breve s_i'(r_i)$ may not be invertible. In this case, $r_i$ can be any solution that satisfies $\breve s_i'(r_i) = \beta$ and $\sum_{i=1}^{N} \pi_i r_i(\beta) = \mathsf r$.}
\end{theorem}

\begin{IEEEproof}
It is easy to check that Eq. (\ref{r_opt}) holds by using the method of Lagrange multipliers.
\end{IEEEproof}

Since $\sum_{i=1}^{N} \pi_i r_i(\beta)$ monotonically increases with $\beta$, we may adopt a binary search algorithm to find $\beta$ in low complexity. When $s_i(r_i)$ is concave for all $i = 1, \ldots, N$, Problem (\ref{cvx}) reduces to a linear program given by
\begin{equation}
\label{LP}
\begin{array}{rrcl}
\displaystyle \min & \multicolumn{3}{l}{ \sum_{i=1}^{N} \pi_i \alpha_i r_i } \\
\textrm{s.t.} & \sum_{i=1}^{N} \pi_i r_i & =    & \mathsf r   \\
              &                      r_i & \in  & [0,1-q_i],
\end{array}
\end{equation}
the optimal solution of which is given by
\begin{equation}
\label{r_LP}
r_{(i)}^* = \left\{
\begin{array}{cl}
1 - q_{(i)}                  &\textrm{if } i < i^* \\
\pi_{(i^*)}^{-1}\left[\mathsf r - \sum_{i=1}^{i^*-1} \pi_{(i)}\left(1-q_{(i)}\right) \right]                &\textrm{if } i = i^* \\
0                        &\textrm{if } i > i^*,
\end{array}
\right.
\end{equation}
where $(i)$ is the index of $\mathcal F_{(i)}$ with the $i$th largest $\alpha_i$ and $i^* = \max \Big\{ n \Big| \sum_{i=1}^{n} \pi_{(i)}\left(1-q_{(i)}\right) < \mathsf r  \Big\}$.

\subsection{The Optimal Probabilistic Caching with Heterogenous RDI}

Having solved the joint rate-cost allocation problem, we next turn our attention to the optimal probabilistic caching, or more specifically, the maximum caching time $t_i$ of content class $\mathcal F_i$. If $s'_i(1-q_i) \leq \beta$, $t_i$ should be $t_i^{\sup}$ seconds. In this case, a content item of $\mathcal F_i$ with $q_i=0$ and $t_i^{\sup} = \infty$ is always cached until it is requested. If $s'_i(1-q_i) > \beta$, then the hit ratio of $\mathcal F_i$ should be $\varsigma_i (\beta)$. To achieve this hit ratio, we consider two possible scenarios. If $\breve s_i ( \varsigma_i (\beta) ) = s_i ( \varsigma_i (\beta) )$, the maximum caching time is determined by $t_i = P_i^{-1} ( \varsigma_i (\beta) + q_i )$. Otherwise, there must exist a probability $\theta_i \in (0, 1)$ and two rate-cost pairs $(s_i(r_i), r_i)$ and $(s_i(\tilde r_i), \tilde r_i)$, $r_i < \tilde r_i$, satisfying $\theta_i r_i + (1 - \theta_i) \tilde r_i = \varsigma_i (\beta)$. In this case, we randomly set the maximum caching time to be $P_i^{-1} ( r_i + q_i )$ with probability $\theta_i$ and to be $P_i^{-1} ( \tilde r_i + q_i )$ with probability $(1-\theta_i)$. When $r_i=0$, the user does not cache the content item with probability $\theta_i$.

\subsection{Arithmetic Caching}

In this subsection, we are interested in a practical situation, in which a user may not have any global knowledge about the content arrival processes, namely, $\lambda$ and $\pi_i$. Therefore, it is not possible for the user to formulate a joint rate-cost allocation problem (\ref{cvx}). Fortunately, the structural result in Theorem \ref{solution} implies an arithmetic caching method without any need of $\lambda$ and $\pi_i$.

The design purpose is to maximize the effective throughput, while limiting the average buffer occupation to be less than or equal to a target value $\mathsf S$. To achieve this goal, the user estimates its average storage cost locally, which is a function of $\beta$, denoted by $\hat S (\beta)$ . If $\hat S (\beta) < \mathsf S$, i.e., the receiver buffer is under-utilized, then we should increase $\beta$ to achieve higher effective throughput.  If $\hat S (\beta) > \mathsf S$, i.e., the receiver buffer is over-utilized, then we should decrease $\beta$ to reduce the storage cost. There are many efficient algorithms to update $\beta$, e.g., $\beta \leftarrow \beta + \left( 1 - \frac{\hat S(\beta)}{\mathsf S} \right) \delta_{\beta}$, where $\delta_{\beta}$ is the step size. However, how to increase the convergence speed is beyond the scope of this paper. Given $\beta$, the user will decide whether and how long a content file should be cached based on its RDI $p_i(x)$, which can be estimated according to its key words or labels. In other words, the user is capable of harvesting storage-efficient content files in air.

\subsection{Communication-Storage Tradeoff with heterogeneous RDI}

Given a target hit ratio $\mathsf r \in \left[0, \sum_{i=1}^{N} \pi_i (1-q_i)\right)$, we may obtain the minimum storage cost, or the mean service time, $s^*(\mathsf r)$ by solving problem (\ref{cvx}). As a result, the overall rate-cost function $s^*(\mathsf r)$ characterizes the optimal communication-storage tradeoff with heterogenous RDI.

Let $\breve r (s)$ denote the overall cost-rate function, which is the inverse function of $s^*(\mathsf r)$. In other words, $\breve r (s)$ presents the maximum hit ratio given a mean caching time $s$. The domain of $\breve r (s)$ is given by $s \in [0,s^{\sup}]$, where $s^{\sup} = \sum_{i=1}^{N} \pi_i (\nu_i + q_i t_i^{\sup})$. Furthermore, $\breve r (s)$ is bounded by $0 \leq \breve r (s) \leq 1$. Similar to the proof of Lemma \ref{lem_convex}, we may show their concavity and convexity in the following.

\begin{lemma}
\label{lem_overall}
$\breve r (s)$ and  $s^*(\mathsf r)$ are increasing concave and convex functions, respectively.
\end{lemma}

Noting that a convex function can be uniformly approximated by a $C^{\infty}$-function, we may assume that $\breve r (s)$ and $s^*(\mathsf r)$ are differentiable. Without solving problem (\ref{cvx}), the structural result (\ref{r_opt}) implies some useful properties of $\breve r'(s)$, as shown in the following lemma.

\begin{lemma}
\label{lem_der_overall}
The first order derivative of $\breve r(s)$ for $s=0$ and $s = s^{\sup}$ can be determined by $\breve r' (0) = \underset{1 \leq i \leq N}{\max}  \breve r'_i (0)$ and $\breve r' (s^{\sup}) = \underset{1 \leq i \leq N}{\min}  \breve r'_i (s^{\sup})$, respectively. When $s_i(r)$ are concave for all $i$, $\breve r'(0) = \alpha_{(1)}^{-1}$ and $\breve r' (s^{\sup}) = \alpha_{(N)}^{-1}$. When $s_i(r)$ are convex for all $i$, $\breve r'(0) = \underset {1 \leq i \leq N}{\max} p_i(0)$ and $\breve r'(0) = \underset {1 \leq i \leq N}{\min} \frac{p_i\left(t_i^{\sup}\right)}{q_i}$.
\end{lemma}

\subsection{A Unified Framework of Probabilistic Caching}

It is worth presenting a unified framework of probabilistic caching. Given the $p.d.f.$s of the maximum caching time $t_i$ of content class $\mathcal F_i$, $f_i(x)$, the hit ratio of content class $\mathcal F_i$ is given by $\int_0^{\infty} P_i(x) f_i (x) d x -q_i$ from the law of total probability. As a result, the overall hit ratio is presented by
\begin{equation}
\label{hit_rate_prob}
r(f_1,\ldots,f_N) = \sum_{i=1}^{N} \pi_i \int_0^{\infty} P_i(x) f_i (x) d x -q,
\end{equation}
where $q=\sum_{i=1}^{N} \pi_i q_i$ means the overall undemand probability. From Eq. (\ref{waiting}) that represents the random caching time for a given $t_i$, the $p.d.f.$ of the caching time of content class $\mathcal F_i$ is determined by $g_i(x) = p_i(x) + [q_i +1 - P_i(x)]f_i(x)$. Therefore, the $c.d.f.$ and expectation of the caching time are presented by
\begin{eqnarray}
\label{cdf_W_prob}
G(x) & = & \sum_{i=1}^{N}\pi_i \left[ P_i(x) + (q_i +1)F_i(x) - \int_0^{x} P_i(y) f_i (y) d y \right], \\
\label{mean_W_prob}
s(f_1,\ldots,f_N) & = & \sum_{i=1}^{N}\pi_i \left[\nu_i + (q_i + 1) \mathbbm E \{t_i\} - \int_0^{\infty} x P_i(x) f_i (x) d x \right].
\end{eqnarray}

A careful reader may see that the minimization of the storage cost $s(f_1,\ldots,f_N)$ in Eq. (\ref{mean_W_prob}) subject to a hit ratio constraint on $r(f_1,\ldots,f_N)$ in Eq. (\ref{hit_rate_prob}) is a variational problem, because $f_1(x),\ldots,f_N(x)$ are probability density functions. Fortunately, it is equivalent to the joint rate-cost allocation problem (\ref{cvx}), which is a convex optimization. Then the solution to problem (\ref{cvx}), namely, $r_i^*(\beta)$ given by Eq. (\ref{r_opt}), allow us to infer $f_i(x)$ as shown in subsection IV-B. This explains why we focus on the static caching along with its probabilistic time-sharing policy first in Section III. However, the unified framework of probabilistic caching is useful in the next section where caching with finite buffer are considered.

\section{Caching with Finite Buffer}
In this section, we investigate caching with finite buffer, the size of which is denoted by $L$. By buffer size $L$, we mean the buffer may cache at most $L$ content items.\footnote{Throughout this section, we assume that all the content items have the same file size, i.e., $b_k = B$, for all $k$, and the buffer may cache at most $LB$ bits.} With a finite buffer, a content file cannot be cached when the buffer is full. Since the dropped content files have no contribution to the effective throughput, we have $R = \lambda B r (1-\epsilon)$, where $\epsilon$ denote the blocking probability of content items due to buffer overflow. As a result, we shall first present the blocking probability $\epsilon$. Then the effective throughput $R$ is maximized under the buffer size constraint. It is interesting to see that the cost-rate function $\breve r(s)$ of caching with infinite buffer also plays a key role in the finite buffer caching system.

\begin{figure}[!t]
\centering
\includegraphics[width=6in]{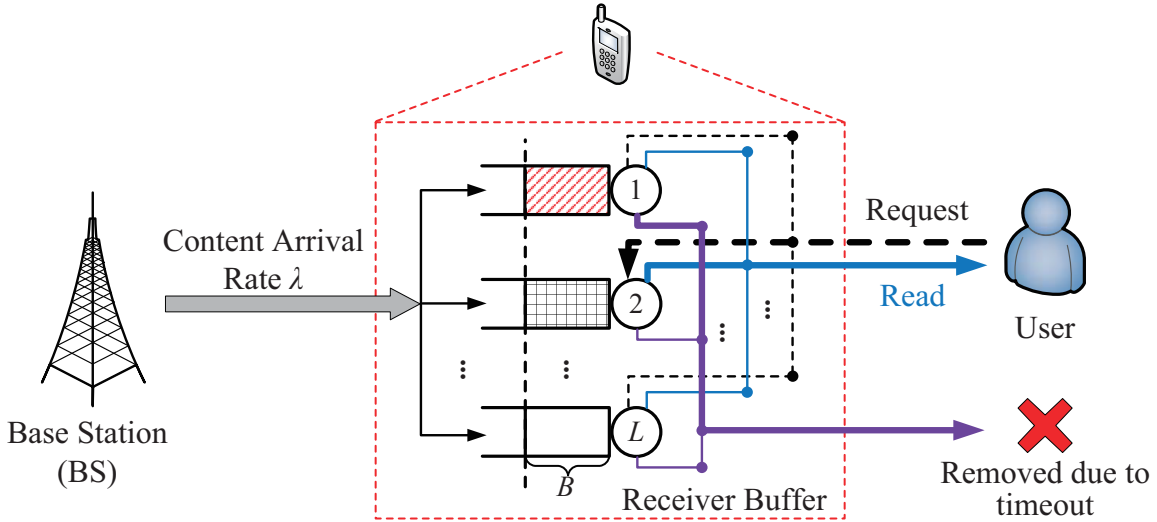}
\caption{A $G/GI/L/0$ model of finite buffer.}
\label{fig_queue_finite}
\end{figure}

\subsection{Blocking Probability of Content Items}

Using Kendall's notation, we formulate a $G/GI/L/0$ queue model\footnote{The queueing model of Lemma \ref{lem_S} is not a $G/GI/L/0$ queue because customers may occupy different sizes of buffer space. Fortunately, the proof of Lemma \ref{lem_S} only relies on Little's Law, that holds for arbitrary queues.} of the buffer state in order to derive $\epsilon$. Each content item is regarded as a customer, the interarrival times of which, $Z_k = \tau_k - \tau_{k-1}$, have a general distribution. The customer's time spent in the equivalent system is equal to the caching time $W$ of the corresponding content file, which is independent of the queue state. As a result, once a customer joins the queue, it is served immediately by one of the parallel servers. Its service time is also equal to the the corresponding content item's caching time. Furthermore, there are $L$ parallel servers in total given the buffer size $L$. If all $L$ servers are busy when a content file arrives, it has to be dropped.

The $G/GI/L/0$ model, and especially its special case $M/GI/L/0$ model, have been extensively studied in the call admission control (CAC) of classic circuit switched networks. In the $M/GI/L/0$ model, the interarrival times $Z_k$ are independent and obey the exponential distribution. Its accurate blocking probability is given by the Erlang B formula, i.e., $\epsilon(s) = B(L,\lambda s)$, which is defined as
\begin{equation}
\label{Erlang_B}
B(L,\lambda s) = \frac{\frac{(\lambda s)^L}{L!}}{\sum_{l=0}^{L} \frac{(\lambda s)^l}{l!}},
\end{equation}
where $s$ is the mean service time, or equivalently, the mean caching time.

Next, we turn our attention to the more general $G/GI/L/0$ model. A diffusion approximation provides a very well approximated blocking probability in the heavy traffic scenario with large values of $L$ and $\lambda s$. From \cite{Whitt}, we have $\epsilon(s) \approx \tilde B (L,\lambda s,z_G)$, which is defined as
\begin{equation}
\label{Whitt_Approx}
\tilde B(L,\lambda s,z_G)=
\sqrt{\frac{z_G}{\lambda s}}
\frac{ \phi\left( \frac{L-\lambda s}{\sqrt{\lambda s z_G}} \right) } { \Phi\left( \frac{L-\lambda s}{\sqrt{\lambda s z_G}} \right) },
\end{equation}
where $\phi(x)$ and $\Phi(x)$ denote the standard normal probability and cumulative distribution functions, respectively, and $z_G$ denotes the asymptotic peakedness of the arrival process $A(\tau)$ with respect to the $c.d.f.$ of the caching time, $G(x)$. More particularly, the asymptotic peakedness is given by
\begin{equation}
\label{peakedness}
z_G = 1+ \frac{c^2 -1}{s} \int_{0}^{\infty} [1-G(x)]^2 dx,
\end{equation}
where $c^2 = \underset{\tau \to \infty} {\lim} \frac{\mathrm {Var} [A(\tau)] }{\lambda \tau}$, as shown by Borovkov in \cite{Borovkov}. For a renewal process $A(\tau)$, we have $c^2=\sigma^2_Z$. In contrast to the $M/GI/L/0$ model, the blocking probability of the $G/GI/L/0$ model is determined by $G(x)$ in Eq. (\ref{cdf_W_prob}), rather than the mean caching time $s$ only. Furthermore, it is worth noting that Eqs. (\ref{Erlang_B}) and (\ref{Whitt_Approx}) are consistent with each other for Poisson arrival, because the $M/GI/L/0$ model is a special case of the $G/GI/L/0$ model. More specifically, we have $c^2=1$ and $z_G = 1$ in the $M/GI/L/0$ model. As a result, Eq. (\ref{Whitt_Approx}) is reduced to $\tilde B(L,\lambda s,1)$, which is approximately equal to $B(L,\lambda s)$ according to Hayward approximation \cite{Whitt}.

Having obtained the blocking probability $\epsilon(s)$, we next present some useful bound and approximation to be adopted in the next subsection.
\begin{lemma}
\label{lem_epsilon_bound}
For large values of $L$ and $\lambda s$, the blocking probability is upper bounded by
\begin{equation}
\label{epsilon_bound}
\epsilon(s) \leq B(uL,u\lambda s),
\end{equation}
where $u = \frac{1}{c^2 \vee 1} \in (0,1]$.
\end{lemma}

\begin{IEEEproof}
Our proof starts with an observation from Eq. (\ref{Whitt_Approx}) that $\tilde B (L,\lambda s,z_G) = \tilde B\left(\frac{L}{z_G},\frac{\lambda s}{z_G},1\right) $. According to Hayward approximation \cite{Whitt}, we have $\tilde B\left(\frac{L}{z_G},\frac{\lambda s}{z_G},1\right) \approx B\left(\frac{L}{z_G},\frac{\lambda s}{z_G}\right)$. By recalling that $s=\int_{0}^{\infty} [1-G(x)] dx$, we have $0 < \frac{1}{s} \int_{0}^{\infty} [1-G(x)]^2 dx \leq 1$ in Eq. (\ref{peakedness}). Therefore, $z_G$ is upper bounded by $z_G \leq u^{-1}$. Since $B\left(\frac{L}{z_G},\frac{\lambda s}{z_G}\right)$ is an increasing function of $z_G$, Eq. (\ref{epsilon_bound}) holds.
\end{IEEEproof}

Lemma \ref{lem_epsilon_bound} implies that the blocking provability of a $G/GI/L/0$ queue with arrival rate $\lambda$ is upper bounded by that of an $M/GI/uL/0$ queue with arrival rate $u\lambda$. More importantly, The upper bound in Eq. (\ref{epsilon_bound}) relies on the mean caching time $s$ only, rather than the $c.d.f.$ of the caching time $G(x)$. For large $L$ and $\lambda s$, Eq. (\ref{epsilon_bound}) also provides a well approximated blocking probability due to the scaling property of Erlang B formula. Alternatively, when $L < \lambda s$, there exists a more simple approximation of $\epsilon(s)$, as shown in the following lemma.

\begin{lemma}
\label{lem_epsilon_approx}
For $\frac{\lambda s}{L} > 1$, the blocking probability is approximated by
\begin{equation}
\label{epsilon_approx}
\epsilon(s) \approx 1 - \frac{L}{\lambda s}.
\end{equation}
\end{lemma}
\begin{IEEEproof}
See Eq. (26) in subsection 6.3 of \cite{Whitt}.
\end{IEEEproof}

\subsection{Effective Throughput of Caching with Finite Buffer}
In this subsection, we shall maximize the effective throughput of caching with finite buffer. Recalling that $R = \lambda B r (1-\epsilon)$, we present the effective throughput maximization problem for the general $G/GI/L/0$ system as
\begin{equation}
\label{opt_G_finite}
\underset{f_1(x),\ldots,f_N(x)}{\max}
\lambda B r(f_1,\ldots,f_N) \left[ 1 - \tilde B \left(L,\lambda s(f_1,\ldots,f_N), z_G(f_1,\ldots,f_N) \right) \right],
\end{equation}
in which $r(f_1,\ldots,f_N)$, $s(f_1,\ldots,f_N)$, and $z_G(f_1,\ldots,f_N)$ are determined by Eqs. (\ref{hit_rate_prob}), (\ref{mean_W_prob}), and (\ref{peakedness}), and $f_i(x)$ are $p.d.f.$s of $t_i$. As a result, Problem (\ref{opt_G_finite}) is a variational problem that is hard to solve. To overcome this, we present an alternative optimization problem given by
\begin{equation}
\label{r_finite}
\underset{0 \leq s \leq s^{\sup}}{\max}
r(L,s) = \breve r (s) [1 - B(uL,u\lambda s) ],
\end{equation}
where $\breve r (s)$, defined in subsection VI-B, is the overall cost-rate function of the infinite buffer system, namely, the inverse function of $s^*(\mathsf r)$.

Before solving Problem (\ref{r_finite}), we first illustrate that it optimizes both the $M/GI/L/0$ and $G/GI/L/0$ systems effectively. For the $M/GI/L/0$ system, the objective function of (\ref{r_finite}) represents exactly the maximal achievable hit ratio for given $s$ and $L$. This is simply due to the following two facts. First, $\breve r (s)$ is maximum hit ratio of the infinite buffer system for given $s$. Second, since $u=1$ for Poisson arrivals, $B(uL,u\lambda s)$ gives the accurate blocking probability. For $G/GI/L/0$ system, the objective function of (\ref{r_finite}) represents an achievable lower bound of the hit ratio given $s$ and $L$, due to the fact that $\epsilon(s) \leq B(uL,u\lambda s)$ in Lemma \ref{lem_epsilon_bound}. Furthermore, this lower bound approximates the hit ratio well for large $\lambda s$ and $L$, thanks to the scaling property of the Erlang B formula. In summary, Problem (\ref{r_finite}) maximizes not only the accurate hit ratio of the $M/GI/L/0$ system, but also the lower bound of the hit ratio of the $G/GI/L/0$ system.

Having justified Problem (\ref{r_finite}), we next solve it. Since it is a one-dimensional optimization, the optimal mean caching time $s^* = \underset{0 \leq s \leq s^{\sup}}{\arg \max} r(L,s)$ can be numerically found in low complexity. In the appendix, we shall further present the quasi-convexity of this problem, which leads to a binary search algorithm with lower complexity. To shed more lights on $s^*$, we present the following two lemmas.

\begin{lemma}
\label{lem_s_opt}
The optimal mean caching time $s^*$ is either $s^* = s^{\sup}$ or a solution to
\begin{equation}
\label{s_opt}
\breve r'(s) -
\left[\frac{1}{1 - B(uL,u\lambda s)} \frac{L}{s} - u \lambda \right]
B(uL,u\lambda s) \breve r(s) =0.
\end{equation}
\end{lemma}

\begin{IEEEproof}
Our proof relies on the extreme value theorem \cite{Zorich}. We first notice that $s^* \neq 0$ because $r(L,0) = 0$. Thus, $s^*$ is either $s^{\sup}$ or a solution to $r'(L,s) = 0$. By noting that $B'(uL,u\lambda s) = u\lambda B(uL,u\lambda s)
\left[\frac{L}{u\lambda s} - 1 + B(uL,u\lambda s)\right]$, we have
\begin{equation}
\label{dr_finite}
r'(L,s) = \left[1 - B(uL,u\lambda s) \right]
\left\{
\breve r'(s) - \left[\frac{1}{1 - B(uL,u\lambda s)} \frac{L}{s} - u \lambda \right] B(uL,u\lambda s) \breve r(s)
\right\}.
\end{equation}
Recalling $B(uL,u\lambda s) < 1$, we may conclude that $r'(L,s) = 0$ if and only if Eq. (\ref{s_opt}) holds.
\end{IEEEproof}

\begin{lemma}
\label{lem_s_opt_finite}
The optimal mean caching time must be finite, i.e., $s^* < \infty$, even if $s^{\sup} = \infty$.
\end{lemma}

\begin{IEEEproof}
Due to the fact that $s^* \leq s^{\sup}$, we have $s^* < \infty$ if $s^{\sup} < \infty$. Otherwise, let us note that $r(L,\infty) = 0$. As a result, we get $s^* < \infty$.
\end{IEEEproof}

Lemma \ref{lem_s_opt_finite} implies that a finite maximum caching time must be set for content class $\mathcal F_i$, unless it satisfies $\nu_i + q_i t_i^{\sup} < \infty$. Although Lemmas \ref{lem_s_opt} and \ref{lem_s_opt_finite} cannot give an analytical solution of $s^*$ in general, they are adopted to present the optimal caching policies in the scenario with small arrival rate, as well as, the scenarios with large and small buffers.

\begin{theorem}
\label{thm_small_lambda}
When $\lambda \leq \breve r'(s^{\sup})$, we have $s^* = s^{\sup}$.
\end{theorem}

\begin{IEEEproof}
Due to the fact that $\breve r(s) \leq 1$ and $u\lambda>0$, Eq. (\ref{dr_finite}) can be lower bounded by $r'(L,s) \geq \left[1 - B(uL,u\lambda s) \right]
\left[ \breve r'(s) - \frac{B(uL,u\lambda s)}{1 - B(uL,u\lambda s)} \frac{L}{s} \right]$. Moreover, it is not difficult to see that $\breve r'(s)$ and $\frac{B(uL,u\lambda s)}{1 - B(uL,u\lambda s)}\frac{L}{s}$ are decreasing and increasing functions respectively for $0 \leq s \leq s^{\sup}$. As a result, we get $r'(L,s) \geq \left[1 - B(uL,u\lambda s) \right]
\left[ \breve r'(s^{\sup}) - \frac{B(uL,u\lambda s^{\sup})}{1 - B(uL,u\lambda s^{\sup})} \frac{L}{s^{\sup}} \right]$. In other words, $r'(L,s)>0$ if
\begin{equation}
\label{s_sup_ineq}
\breve r'(s^{\sup}) \geq \frac{L}{s^{\sup}} \frac{B(uL,u\lambda s^{\sup})}{1 - B(uL,u\lambda s^{\sup})}.
\end{equation}

By substituting Eq. (\ref{Erlang_B}) into Eq. (\ref{s_sup_ineq}) and noting that $\frac{(u \lambda s^{\sup})^n}{n!} < \sum_{l=0}^{uL-1} \frac{(u \lambda s^{\sup})^l}{l!} $ for all $n = 0, \ldots, uL-1$, we see Eq. (\ref{s_sup_ineq}) holds if there exists an integer $0 \leq n \leq uL-1$ such that
\begin{equation}
\label{Er_B_ineq}
\breve r'(s^{\sup}) \geq \frac{L}{s^{\sup}} \frac{n!}{(uL)!}(u \lambda s^{\sup})^{uL - n}.
\end{equation}
By inserting $n = uL-1$ into Eq. (\ref{Er_B_ineq}), we obtain $\lambda \leq \breve r'(s^{\sup})$. In this case, we have $r'(L,s)>0$ for $0 \leq s \leq s^{\sup}$, and thus $s^* = s^{\sup}$.
\end{IEEEproof}

Theorem \ref{thm_small_lambda} shows that $s^* = s^{\sup}$ if the arrival rate $\lambda$ is relatively small. The arrival rate threshold can be determined by Lemma \ref{lem_der_overall} without solving Problem (\ref{cvx}). A careful reader may notice that Eq. (\ref{Er_B_ineq}) also leads to a series of sufficient conditions for $s^* = s^{\sup}$. The following theorem gives two buffer size thresholds, beyond which the optimal mean caching time achieves its upper bound, i.e., $s^* = s^{\sup}$. In other words, it presents the optimal caching policy for the large buffer scenario.

\begin{theorem}
\label{thm_large_buffer}
When the buffer size is large, i.e.,
\begin{equation}
\label{L_th_smax}
L \geq \min \left\{ \frac{\lambda^2 s^{\sup}}{\breve r'(s^{\sup})} + \frac{1}{u}, \max \left\{ \frac{1}{2u} \ln \frac{\lambda e^2 }{us^{\sup}} - \frac{1}{u} \ln \breve r'(s^{\sup}), \lambda s^{\sup} e^2 \right\} \right\},
\end{equation}
we have $s^* = s^{\sup}$.
\end{theorem}

\begin{IEEEproof}
By substituting $n = uL-2$ into Eq. (\ref{Er_B_ineq}), we get $\breve r'(s^{\sup}) \geq \frac{u \lambda^2 s^{\sup}}{uL-1}$, that yields $L \geq \frac{\lambda^2 s^{\sup}}{\breve r'(s^{\sup})} + \frac{1}{u}$. When $n = 0$, Eq. (\ref{Er_B_ineq}) reduces to $\breve r'(s^{\sup}) \geq \frac{L}{s^{\sup}} \frac{(u \lambda s^{\sup})^{uL}}{(uL)!}$. By recalling Stirling's formula $(uL)! \geq \sqrt{2 \pi} (uL)^{uL+\frac{1}{2}} e^{-uL}$ and $0 < u \leq 1$, we see that Eq. (\ref{Er_B_ineq}) holds if $\breve r'(s^{\sup}) \geq \sqrt{\frac{\lambda e}{us^{\sup}}} \left( \frac{\lambda s^{\sup}e}{L} \right) ^{uL-\frac{1}{2}}$, which implies $L \geq \max \left\{ \frac{1}{2u} \ln \frac{\lambda e^2 }{us^{\sup}} - \frac{1}{u} \ln \breve r'(s^{\sup}), \lambda s^{\sup} e^2 \right\}$.
\end{IEEEproof}

From Theorems \ref{thm_small_lambda} and \ref{thm_large_buffer}, we have $s^* = s^{\sup}$ for the large buffer and small arrival rate scenarios. When $s^* = s^{\sup}$, the maximal caching time $t_i^*$ should also achieve its upper bound, i.e., $t_i^* = t_i^{\sup}$ for all $i$. In other words, a cached content item will be removed unless it is requested or will never be requested. As a result, if a content item is not blocked, its hit ratio is equal to its demand probability. In this case, the hit ratio is obtained by $[1 - B(uL,u\lambda s)]\sum_{i=1}^{N} \pi_i (1-q_i)$. Since $B(uL,u\lambda s) \approx 0$ for large $L$, the hit ratio and effective throughput are respectively presented by
\begin{eqnarray}
\label{r_large_buffer}
r_1^*(q) & \approx & 1 - q, \\
\label{R_large_buffer}
R_1^*(q) & \approx & \lambda B (1-q).
\end{eqnarray}
From Eq. (\ref{R_large_buffer}), the effective throughput is a linear increasing function of the demand probability $(1-q)$ and the content data rate $\lambda B$, but is irrelevant to the buffer size $L$. In other words, the effective throughput is limited by the demand probability for small $\lambda$ or large $L$. Therefore, we shall refer to this scenario as the demand probability limited region. In this case, caching with finite buffer is approximately equivalent to that with infinite buffer.

Next we turn our attention to the small buffer scenario. Based on Lemma \ref{lem_epsilon_approx}, we obtain an approximate but analytical result of $s^*$, as shown in the following theorem.

\begin{theorem}
\label{thm_small_buffer}
When $L \ll \lambda$, we have $s^* \approx \frac{L}{\lambda}$.
\end{theorem}

\begin{IEEEproof}
Our proof relies on the observation that $L < \lambda s$ for $s > s^*$. In this case, the blocking probability has a good approximation Eq. (\ref{epsilon_approx}) as shown in Lemma \ref{lem_epsilon_approx}. By substituting Eq. (\ref{epsilon_approx}) into Eq. (\ref{r_finite}), we get
\begin{equation}
\label{r_small_L}
r(L, s) \approx \frac{L}{\lambda}\frac{\breve r (s)}{s}.
\end{equation}
Since $\breve r (s)$ is convex and $\breve r (0) = 0$, Eq. (\ref{r_small_L}) is a decreasing function of $s$. Recalling that $L \ll \lambda$, we have $\frac{L}{\lambda} \approx 0$ and hence $r\left(L, \frac{L}{\lambda}\right)$ is approximately the supremum of the hit ratio.
\end{IEEEproof}

Let us substitute $\breve r' (0) = \underset{s \to 0}{\lim} \frac{\breve r (s)}{s}$ into Eq. (\ref{r_small_L}). Then the hit ratio and effective throughput in the small buffer scenario are respectively given by
\begin{eqnarray}
\label{r_small_buffer}
r_2^*(L) & = & \frac{L}{\lambda} \breve r' (0), \\
\label{R_small_buffer}
R_2^*(L) & = & LB \breve r' (0),
\end{eqnarray}
where $\breve r' (0)$ can be determined by Lemma \ref{lem_der_overall} without solving Problem (\ref{cvx}).

From Eq. (\ref{R_small_buffer}), the effective throughput is a linear increasing function of the buffer size $LB$ (bits), but is irrelevant to the arrival rate $\lambda$. In other words, the effective throughput is limited by the buffer size when $L < \lambda s$. Therefore, we shall also refer to the scenario that the number of content items arrived in the mean caching time is greater than the buffer size as the buffer limited region. In this case, only the content items with the maximal $\breve r'_i (0)$ can be cached because $\breve r' (0) = \underset{1 \leq i \leq N}{\max}  \breve r'_i (0)$.

Finally, Theorems \ref{thm_small_lambda}, \ref{thm_large_buffer}, and \ref{thm_small_buffer} are consistent with each other. To demonstrate this, we first focus on the scenario in which $L \ll \lambda \leq \breve r'(s^{\sup})$. Recalling $L \geq 1$, we see $\breve r'(s^{\sup}) \geq \lambda \gg 1$. Due to the convexity of $\breve r(s)$ and the fact that $\breve r(s^{\sup}) \leq 1$, we also have $\breve r'(s^{\sup}) \leq \frac{1}{s^{\sup}}$. As a result, $s^{\sup}$ must be very small. In other words, any $s^* \in (0, s^{\sup}]$ is near optimal,
which validates the consistency between Theorems \ref{thm_small_lambda} and \ref{thm_small_buffer}. When $L \ll \lambda$ and Eqs. (\ref{L_th_smax}) hold simultaneously, we may also show that $s^{\sup} \ll 1$. Similarly, this validates the consistency between Theorems \ref{thm_large_buffer} and \ref{thm_small_buffer}.

\subsection{Arithmetic Caching with Finite Buffer}

In this subsection, we turn our attention to a more practical situation, in which a finite-buffer user has no global knowledge about $\lambda$ and $\pi_i$. Therefore, it is not possible to formulate and solve Problem (\ref{r_finite}). To overcome this, we propose a finite-buffer arithmetic caching policy. Let $\hat R(\beta)$ denote the estimated effective throughput as a function of $\beta$. Via Eq. (\ref{r_opt}), $\beta$ determines the maximum caching time of each content class $\mathcal F_i$. In contrast to the infinite-buffer arithmetic caching, in which $\beta$ is adjusted to meet the average buffer consumption constraint $\hat S (\beta) = \mathsf S$, the finite-buffer arithmetic caching is designed to find $\beta^*$ that maximizes $\hat R(\beta)$. In the appendix, we investigate the quasi-concavity of $r(L,s)$, which results in that $\hat R(\beta)$ increases with $\beta$ when $\beta < \beta^*$ and decreases with the increase of $\beta$ when $\beta > \beta^*$. As a result, the finite-buffer arithmetic caching will first estimate $\hat R\left(2^k\right)$ for $k=1,2,\ldots$, until it finds $k^*$ satisfying $\hat R\left(2^{k^*}\right) > \hat R\left(2^{k^*+1}\right)$. Once $k^*$ is found, $\beta^*$ can be bounded by $0< \beta^* <2^{k^*+1}$. Then binary search methodology will be exploited to find $\beta^*$ in low complexity.

\section{Numerical Results}

In this section, numerical results are presented to validate the theoretical analysis and demonstrate the potential of the proposed storage efficient caching with time-domain buffer reuse. We consider ten content classes, the RDI of which are given by $p_1(x) = \exp (-x) \mathbbm 1_{\{x \geq 0\}}$, $p_2(x) = \mathbbm 1_{\{0 \leq x \leq 1\}}$, $p_3(x) = x \mathbbm 1_{\{0 \leq x \leq 1\}} + (2-x) \mathbbm 1_{\{1 < x \leq 2\}} $, $p_4(x) = x^{-2} \mathbbm 1_{\{x \geq 1\}}$, $p_5(x) = \frac{1}{\pi} \sqrt{\frac{\mathbbm 1_{\{0 < x < 2\}}}{x(2 - x)}} $, $p_6 = \Psi_{0.6}^{3} p_1(x)$, $p_7 = \Psi_{0.5}^{1} p_2(x)$, $p_8 = \Psi_{1}^{2} p_3(x)$, $p_9 = \Psi_{0.4}^{3} p_4(x)$, and $p_{10} = \Psi_{0.2,1}^{4} \Psi_{0.8}^{3} p_5(x)$. Clearly, the first five content classes have RDIs presented in Table \ref{Typical_pdf}, while the others have modified RDIs defined in Table \ref{CP_property}. We assume three content arrival processes, namely, Poisson arrival with $z_G = 1$, deterministic arrival with $c^2 = 0$,\footnote{For deterministic arrival, the queueing system reduces to a $D/GI/L/0$ model.} and phase-type arrival with $c^2 = 2$ that models burst traffics\footnote{The $p.d.f.$ of the content interarrival time $Z$ is given by $f_Z(z)=\frac{5}{3}\exp(-5z) +\frac{40}{3}\exp(-20z)$.}.

\begin{figure}[!t]
\centering
\includegraphics[width=4in,angle=270]{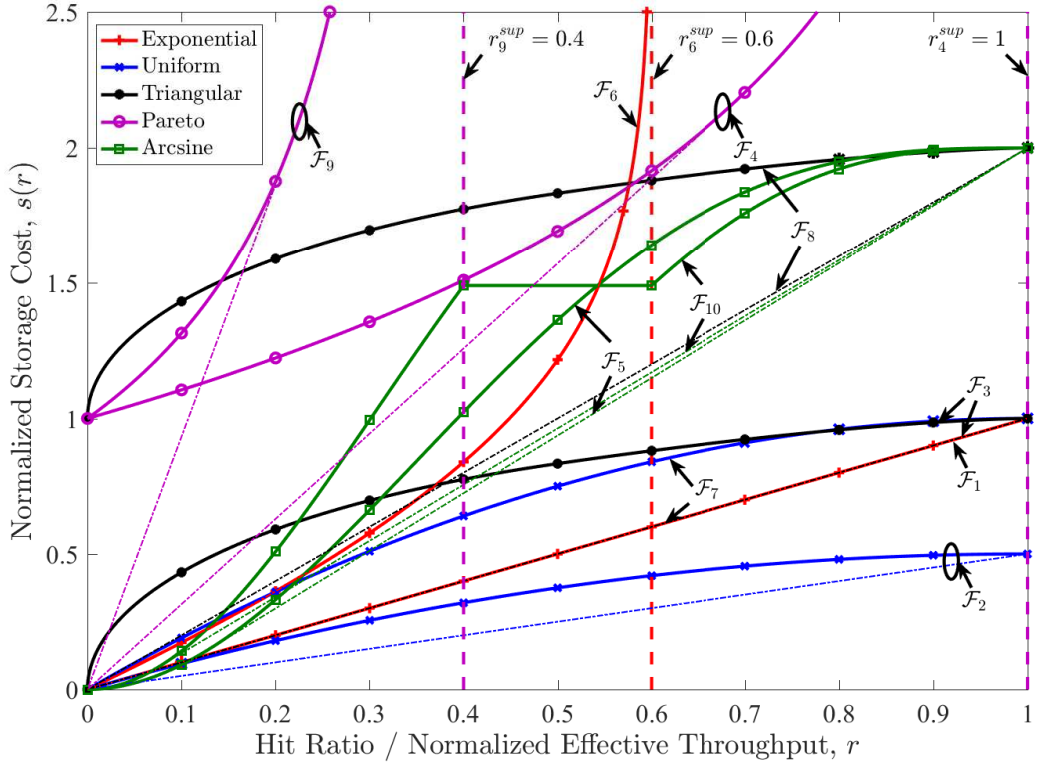}
\caption{The storage cost versus effective throughput curves with infinite buffer and homogeneous RDI.}
\label{fig_inf_homo}
\end{figure}

Fig. \ref{fig_inf_homo} presents both the normalized rate-cost functions $s_i(r)$ and their lower convex envelopes $\breve s_i(r)$ of infinite-buffer caching with homogeneous RDI given by $p_i(x)$, $i = 1, \ldots, 10$. The theoretical results of the rate-cost pairs obtained via Eqs. (\ref{norm_sr}) perfectly match their corresponding simulation results. By comparing $s_i(r)$ and $s_{i+5}(r)$ for $i=1,\ldots,5$, the properties of CP-transform in Table \ref{CP_property} are also validated. A careful reader may notice that only $s_1(r)$ is a linear function, which confirms Corollary \ref{cor_linear_exp}. Moreover, content classes $\mathcal F_i$, $i=2,3,7,8$, with RDIs belonging to uniform and triangular distribution families have convex rate-cost functions. Therefore, their lower envelopes $\breve s_i(r)$ are straight line segments with slopes given by $\alpha_i$. Since $\alpha_1 = \alpha_3 =\alpha_7$, the optimal rate-cost functions $\breve s_i(r)$, $i=1,3,7$, are exactly the same, as shown by a close observation. By contrast, $s_5(r)$ and $s_{10}(r)$ corresponding to Arcsin distribution family are neither convex nor concave. The storage costs of $\mathcal F_i$, $i=4$, $6$, and $9$ are approaching plus infinity when the hit ratio $r_i$ goes to its upper bound $r_4^{\sup}=1$, $r_6^{\sup}=0.6$, and $r_9^{\sup}=0.4$. This is not surprising because $s_i$ is upper bounded by $\mu_i+q_i t_i^{\sup}$, in which we have $\mu_4=\infty$ for Pareto distribution $p_4(x)$, as well as, $t_i^{\sup} = \infty$ and $q_i>0$ for $i=6, 9$. As a result, the storage cost is dominated by the maximum request delay $t_i^{\sup}$ for large $r_i$ and nonzero $q_i$.

\begin{figure}[!t]
\centering
\includegraphics[width=4in,angle=270]{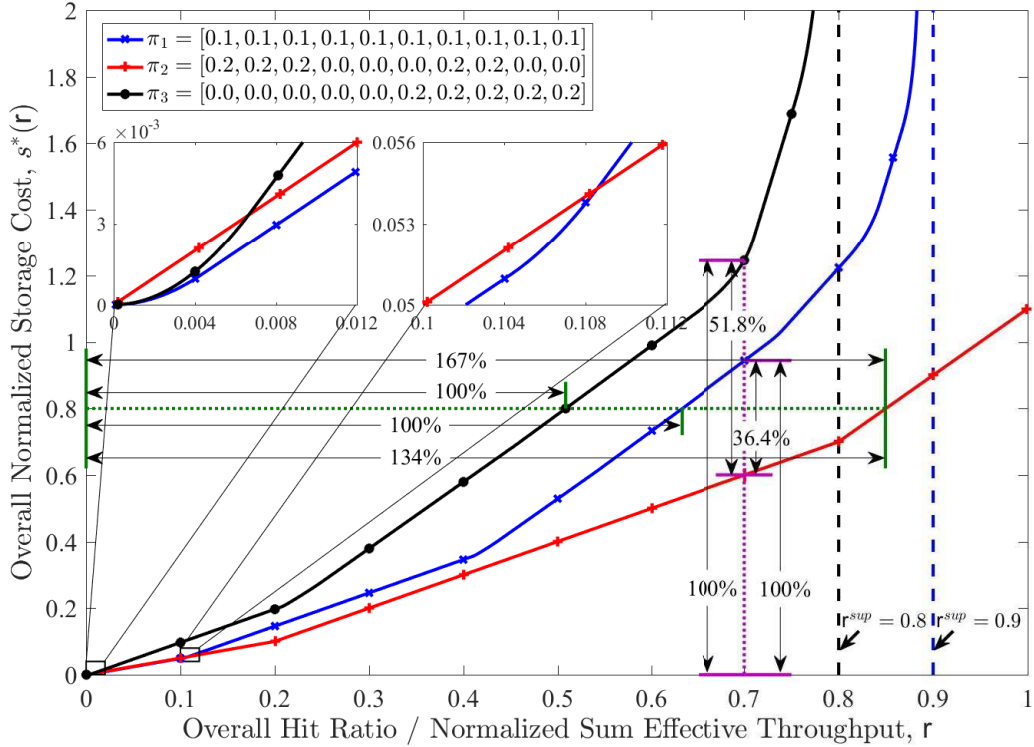}
\caption{The overall storage cost versus normalized sum effective throughput curves with infinite buffer and heterogeneous RDI.}
\label{fig_inf_hete}
\end{figure}

Fig. \ref{fig_inf_hete} presents the overall storage cost versus normalized sum effective throughput curves with infinite buffer and heterogeneous RDI. We denote $\bm{\pi} = [\pi_1, \ldots, \pi_{10}]$ to distinguish three content flows with heterogeneous RDI. More particularly, we consider $\bm{\pi}_1 =[0.1, \ldots, 0.1]$, $\bm{\pi}_2 =[0.2, 0.2, 0.2, 0, 0, 0, 0.2, 0.2, 0, 0]$, and $\bm{\pi}_3 =[0, 0, 0, 0, 0, 0.2, 0.2, 0.2, 0.2, 0.2]$. Their maximum feasible hit ratios are equal to $0.9$, $0.8$, and $1$, respectively. Given a target hit ratio $\mathsf r$, the storage costs are obtained by solving problem (\ref{cvx}). It can be seen that the theoretical results perfectly match their corresponding simulation results. Part of each curve is straight line segment, because there exist content classes with linear rate-cost functions in each content flow. In particular, the overall rate-cost curve of content flow $\bm{\pi}_2$ is piece-wise linear. This is not surprising because $\breve s_i(r)$, $i= 1,2,3,7,8$ are linear. A close observation shows that the slope of the $i$th line segment is equal to the $i$th smallest $\alpha_i$, which validates the optimal solution (\ref{r_LP}). For both flows $\bm{\pi}_1$ and $\bm{\pi}_3$, the overall rate-cost curves go to plus infinity with two vertical asymptotes $\mathsf r = 0.9$ and $\mathsf r = 0.8$, because both flows consist of content classes with $s_i^{\sup} = \infty$, e.g., $\mathcal F_i$, $i=4, 6, 9$. When $\mathsf r_2^{\sup}=1$, the storage cost of flow $\bm{\pi}_2$ is finite, because exponential and Pareto distributed RDIs do not exist.

To provide further insights, we compare the three overall rate-cost curves for large and small $\mathsf r$. When $\mathsf r = 0.7$, the content flow $\bm{\pi}_2$ saves $36.4\%$ and $51.8\%$ storage cost, compared to those of flows $\bm{\pi}_1$ and $\bm{\pi}_3$. When the storage cost is $s = 0.8$, the hit ratio gains of $\bm{\pi}_2$ over $\bm{\pi}_1$ and $\bm{\pi}_3$ are $134\%$ and $167\%$. When $\mathsf r$ is small, however, the content flow $\bm{\pi}_1$ achieves the smallest storage cost. This is not surprising because content flow $\bm{\pi}_1$ consists of $\mathcal F_5$ with the highest storage efficiency, while only the content class with the highest storage efficiency is cached for small $\mathsf r$. The optimality of flow $\bm{\pi}_1$ implies that the storage efficiency may increase with the diversity of candidate content classes. However, when the number of candidate classes is large, it is difficult to estimate the global parameters of the arrival process. As a result, the decentralized arithmetic caching is highly desired in practice.

\begin{figure}[!t]
\centering
\includegraphics[width=4in,angle=270]{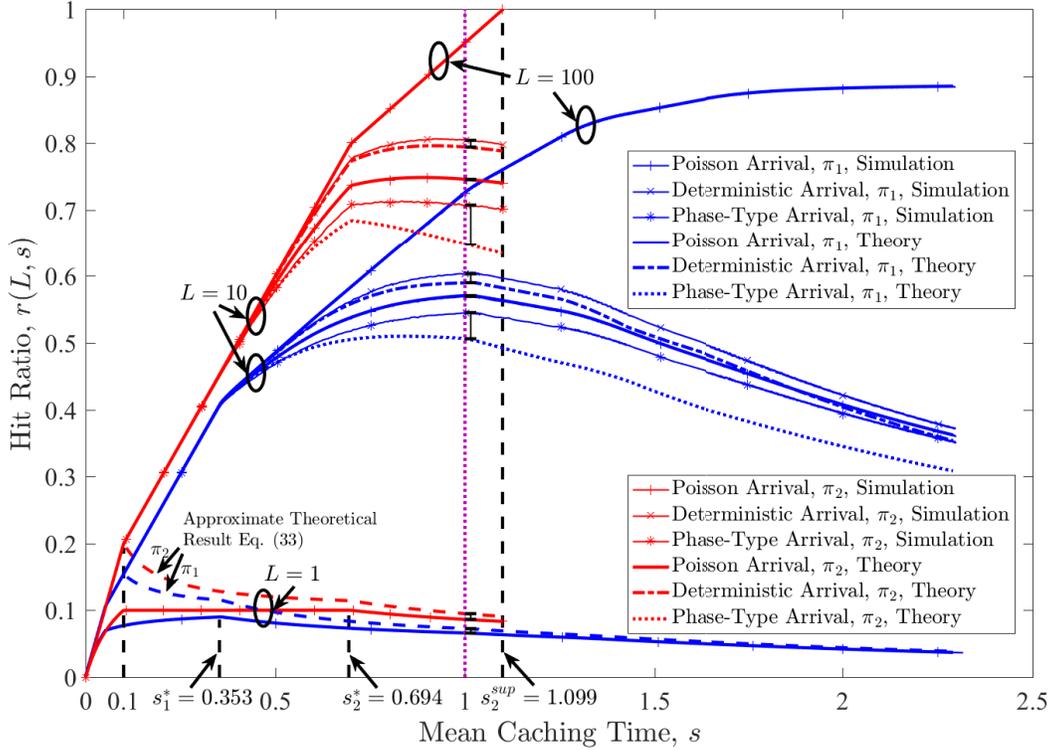}
\caption{The mean caching time versus hit ratio curves with finite buffer.}
\label{fig_finite_rs}
\end{figure}

Next, we turn our attention to the finite buffer scenario. Fig. \ref{fig_finite_rs} presents the mean caching time versus hit ratio curves of content flows $\bm{\pi}_1$ and $\bm{\pi}_2$ with arrival rate $\lambda=10$. We consider three buffer sizes, namely, large size $L=100$, middle size $L=10$, and small size $L=1$. We present hit ratios with Poisson, deterministic, and phase-type arrivals if $L=10$. Otherwise, only Poisson arrival is considered due to the space limitation of Fig. \ref{fig_finite_rs}. For given mean caching time $s$, we solve problem (\ref{cvx}) to determine the probabilistic caching policy, $f_i(t_i)$, based on which the simulation results are obtained. To obtain the theoretical hit ratios, we adopt Erlang B formula Eq. (\ref{Erlang_B}) and diffusion approximation Eq. (\ref{Whitt_Approx}) to calculate the blocking probabilities. Furthermore, the approximate hit ratio given by Eq. (\ref{r_small_L}) is also presented.

It can be seen that the theoretical and simulation results match well with each other. For Poisson arrival, Erlang B formula achieves zero approximation error. When $s=1$, the errors of hit ratios based on approximate blocking probability Eq. (\ref{Whitt_Approx}) are $2.21\%$ and $7.17\%$ for content flow $\bm{\pi}_1$ with deterministic and phase-type arrivals, respectively. Phase-type arrival suffers from the greater approximation error because of its burst nature. Furthermore, the error due to approximation Eq. (\ref{Whitt_Approx}) decreases with the increase of $s$, because the diffusion approximation relies on the heavy traffic assumption. For small buffer $L=1$, Eq. (\ref{r_small_L}) also provides a good approximation of the hit ratio. When $s=1$, the errors are $9.95\%$ and $9.92\%$ for flows $\bm{\pi}_1$ and $\bm{\pi}_2$, respectively. Moreover, the approximation error vanishes with the increase of $s$. A careful reader may further notice that all the curves are quasi-concave and have unique maximum. For $L=100$, the hit ratio increases monotonically with $s$. Therefore, we have $s^* = \infty$ and $s^* = 1.1$ for flows $\bm{\pi}_1$ and $\bm{\pi}_2$, which confirms Theorem \ref{thm_large_buffer}. For $L=1$, the hit ratios of both flows are approximately decreasing functions of $s$ when $s \geq 0.1$. Therefore, we have $s^* \approx 0.1$, which confirms Theorem \ref{thm_small_buffer}.

\begin{figure}[!t]
\centering
\includegraphics[width=4in,angle=270]{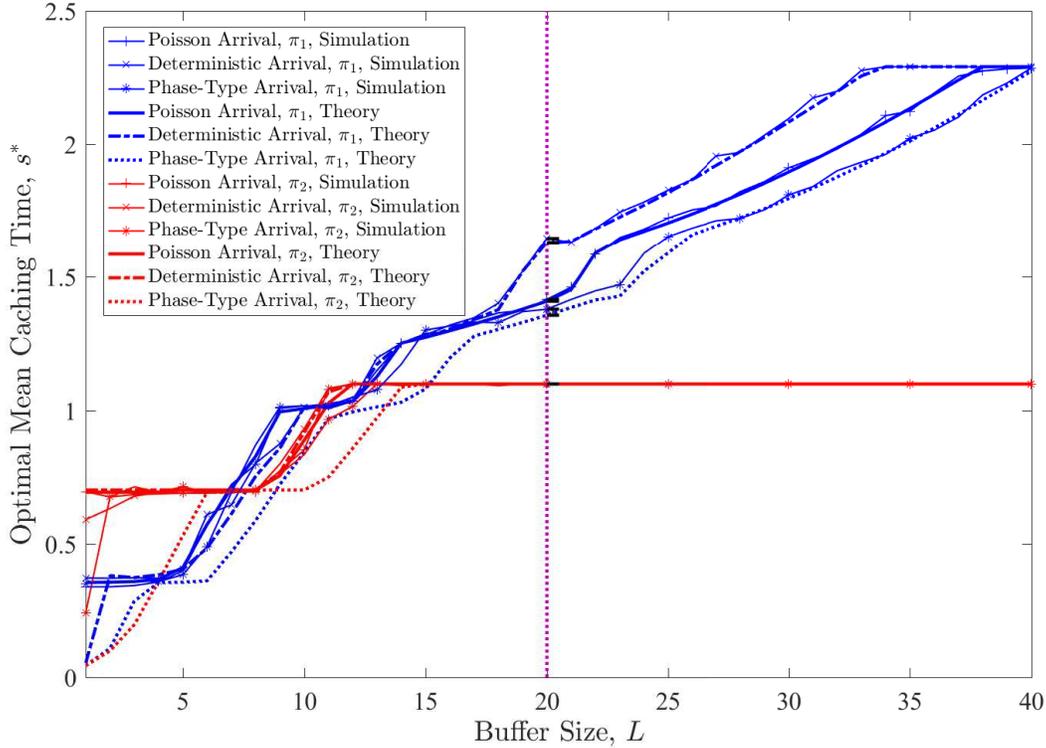}
\caption{The buffer size versus optimal mean caching time curves.}
\label{fig_finite_s_opt_L}
\end{figure}

\begin{figure}[!t]
\centering
\includegraphics[width=4in,angle=270]{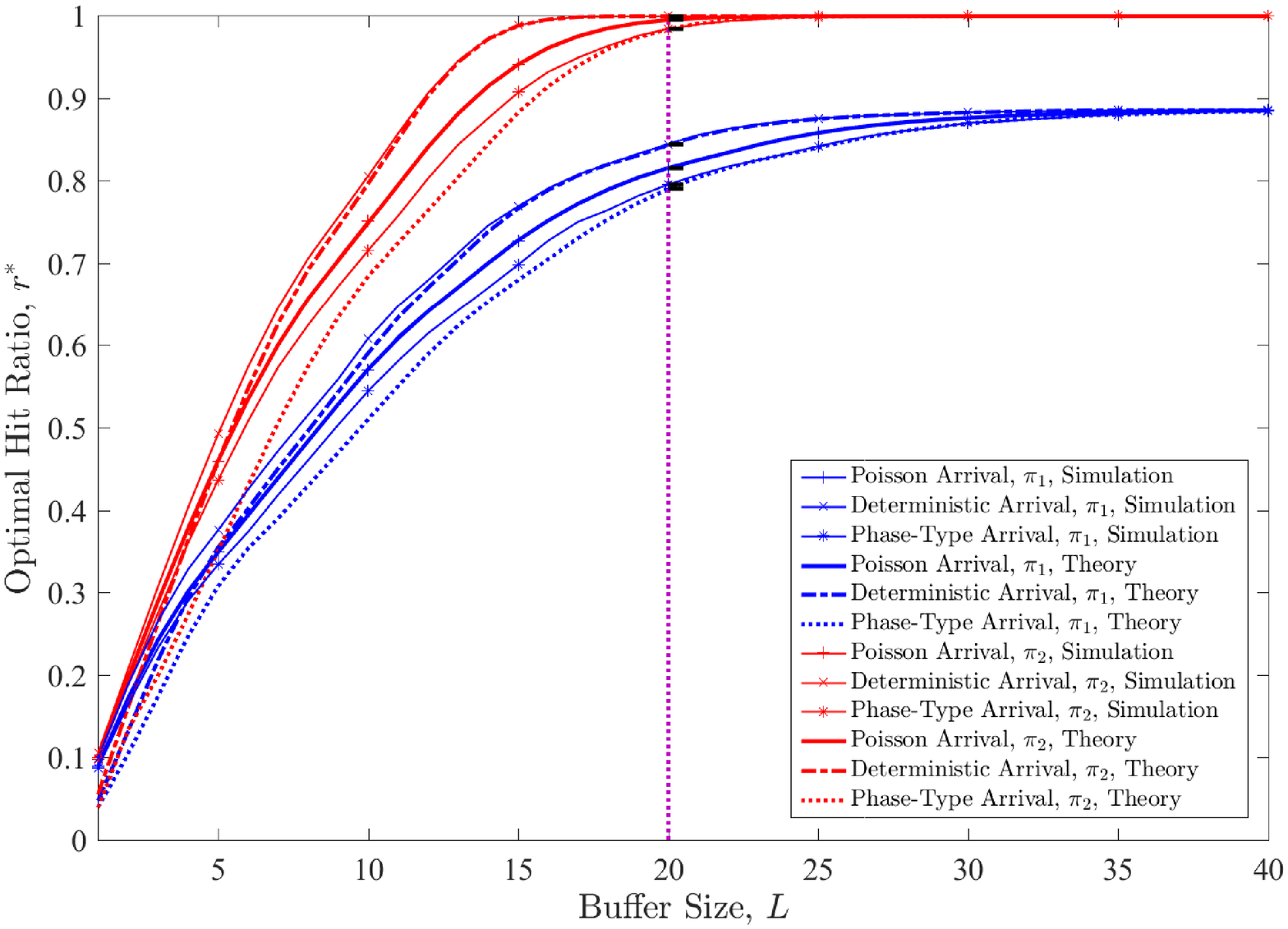}
\caption{The buffer size versus optimal hit ratio curves.}
\label{fig_finite_r_opt_L}
\end{figure}

Figs. \ref{fig_finite_s_opt_L} and \ref{fig_finite_r_opt_L} demonstrate how the buffer size $L$ determines the optimal mean caching time $s^*$ and the optimal hit ratio $r^*$, by presenting $s^*(L)$ and $r^*(L)$ curves, respectively. We consider content flows $\bm{\pi}_1$ and $\bm{\pi}_2$ with three arrival processes, the arrival rates of which are $\lambda = 10$. The theoretical results of $s^*$ and $r^*$ are given by Lemma \ref{lem_s_opt}, in which Eq. (\ref{s_opt}) can be solved numerically. To present the simulation results of $s^*(L)$ and $r^*(L)$, we first adopt the simulation method of Fig. \ref{fig_finite_rs} to plot the $s$ versus $r$ curve, based on which the optimal values $s^*$ and $r^*$ are found numerically. The simulation and theoretical results of $s^*(L)$ and $r^*(L)$ curves match well with each other. Let us focus on the buffer size $L=20$. For flows $\bm{\pi}_1$ and $\bm{\pi}_2$, the errors of $s^*$ are no greater than $1.65\%$ and $0.01\%$, while those of $r^*$ are upper bounded by $0.67\%$ and $0.02\%$. Clearly, these errors are negligible in practice. A careful reader may notice that the simulation curves of $s^*(L)$ are not as smooth as those of $r^*(L)$. This is due to the facts shown in Fig. \ref{fig_finite_rs} that $s^*(L)$ is very sensitive to the simulation errors of $r(L,s)$, while $r^*(L)$ is insensitive to these errors.

To provide further insights, we next focus on the large and small buffer region in Figs. \ref{fig_finite_s_opt_L} and \ref{fig_finite_r_opt_L}. For large $L$, $s^*$ and $r^*$ approach $s^{\sup}$ and the demand probability $1-q$, respectively, which validates Theorem \ref{thm_large_buffer} and Eq. (\ref{r_large_buffer}). In this case, the hit ratio is limited by the demand probability. For flow $\bm{\pi}_1$, $s^*$ always increases with $L$ because $s^{\sup} = \infty$. For flow $\bm{\pi}_2$, $s^*$ remains as $s^{\sup}$ when $L>15$, because $s^{\sup} < \infty$. For small $L$, both $s^*$ and $r^*$ are approximately linear functions of $L$, which validates Theorem \ref{thm_small_buffer} and Eq. (\ref{r_small_buffer}). In this case, the hit ratio is limited by the buffer size.

\begin{figure}[!t]
\centering
\includegraphics[width=4in,angle=270]{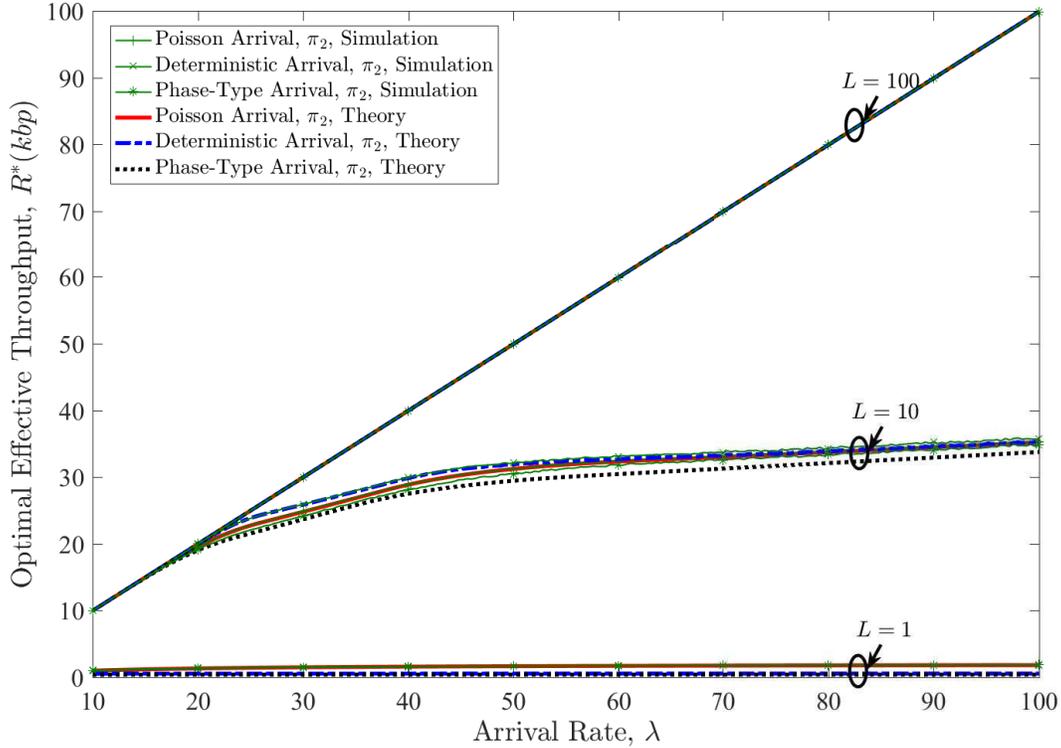}
\caption{The arrival rate versus optimal effective throughput curves with finite buffer.}
\label{fig_finite_R_opt_lambda}
\end{figure}

Finally, Fig. \ref{fig_finite_R_opt_lambda} presents the arrival rate versus optimal effective throughput curves for three buffer sizes, namely, $L=1$, $L=30$, and $L=200$. Due to space limitation, we focus on flow $\bm{\pi}_2$, in which the content file size is set to be $B=1k$bits. For Poisson arrival, the simulation and theoretical results match perfectly with each other. For deterministic and phase-type arrivals, the approximation errors vanishes as the traffic load $\lambda$ increases, because the approximation relies on the heavy traffic assumption. For large buffer $L=200$, the effective throughput increases linearly with the arrival rate. The slopes of the $R^*(\lambda)$ curves are approximately equal to $1$, which confirms Eq. (\ref{R_large_buffer}). For small buffer $L=1$, the effective throughput approximately remains as a constant given by $2kbps$. This observation validates Eq. (\ref{R_small_buffer}) and that the effective throughput is irrelevant to the arrival rate in the buffer limited region. With middle buffer $L=30$, the effective throughput increases linearly with the arrival rate when $\lambda < 20$, but approximately remains unchanged when $\lambda > 60$. From the comparison of $R^*(L)$ curves for three buffer sizes, we see that the buffer size thresholds for both the demand probability and buffer size limited regions increase with the arrival rate, which also confirms Theorems \ref{thm_large_buffer} and \ref{thm_small_buffer}. For a given arrival rate in practice, a user with small buffer tends to cache the content items with the high storage efficiency, while a user with large buffer tends to cache more content items to fully utilize its buffer.

\section{Conclusions}

In this paper, we have investigated storage efficient caching based on time domain buffer sharing in order to strike the optimal communication-storage tradeoff. In particular, we have formulated a queue model, in which Little's law bridges the storage cost and the maximum caching time. By this means, the storage cost has been presented as a function of the hit ratio, also referred to as the rate-cost function. To attain the optimal rate-cost tradeoff, we have conceived probabilistic caching with random maximum caching time. For multiple content classes with different RDIs, the overall storage efficiency has been maximized by solving a joint rate-cost allocation problem. We have also presented storage efficient caching with finite buffer, the effective throughput of which is jointly determined by the hit ratio of caching with infinite buffer and the blocking probability due to buffer overflow. The blocking probability has been derived by using the diffusion approximation or Erlang-B formula, based upon a $G/GI/L/0$ queue along with its special case, $M/GI/L/0$ queue. Then we have presented a one-dimensional quasi-convex optimization to maximize the effective throughput of caching with finite buffer. When the buffer is sufficiently large or small, its approximate but analytical solutions can be found, which significantly simplifies the caching policies. For more practical purposes, we have also presented decentralized arithmetic caching without any need for the global knowledge of content arrival processes. With arithmetic caching, a user may efficiently harvest content files in air by determining whether and how long it should cache a content file based on its local RDI prediction.

\appendix
\section*{Quasi-convexity of Problem (\ref{r_finite})}
In this appendix, we investigate the quasi-convexity of Problem (\ref{r_finite}), which may further reduce the complexity in solving this problem. To test the quasi-concavity of $r(L,s)$ in a unified way, we present a discriminant that is irrelevant to $\breve r (s)$. Based on the discriminant, the quasi-convexity of Problem (\ref{r_finite}) is shown for $L \leq 1000$. The key idea relies on the observation that $r(\tilde L,s)$ is quasi-concave if $ h_L(s) = h(s) [1 - B(L,s)] $ is a quasi-concave function for $L = u \tilde L$ and arbitrary concave increasing function $h(s)$. In this context, we present the following theorem.

\begin{theorem}
\label{thm_QC_D}
For arbitrary concave increasing function $h(s)$, $ h_L(s) = h(s) [1 - B(L,s)] $ is quasi-concave for $s \geq 0$ if $L \leq 5$ or $\Delta_L(l) \geq 0$, for $l=1,2,\ldots,L-5$, where the discriminant $\Delta_L(l)$ is given by
\begin{equation}
\label{QC_D}
\Delta_L(l) = \sum_{i=1}^{L-l-1}\frac{i(i+1)-2L}{(L-i)!(l+i)!}+\frac{2(L-k-1)}{(L-1)!l!}.
\end{equation}
\end{theorem}

\begin{IEEEproof}
From \cite{Convex}, $h_L(s)$ is quasi-concave if and only if $h'_L(s) = 0 \Rightarrow h''_L(s) \leq 0$. Since $h'_L(s) = h'(s)[1 - B(L,s)] - h(s) B'(L,s)$, $h'_L(s) = 0$ means
\begin{equation}
\label{1st_d_0}
h_L(s) = \frac{h'(s)[1 - B(L,s)]}{B'(L,s)}.
\end{equation}
By substituting Eq. (\ref{1st_d_0}) into $h''_L(s) = h''(s)[1 - B(L,s)] - 2h'(s) B'(L,s)-h(s) B''(L,s)$, we get $h''_L(s) = h''(s)[1 - B(L,s)] - h'(s) \left\{ 2B'(L,s) + \frac{[1 - B(L,s)]B''(L,s)}{B'(L,s)} \right\}$. Also note that $h'(s) \geq 0$ and $h''(s) \leq 0$ because $h(s)$ is a concave increasing function. As a result, $h''_L(s) \leq 0$ can be assured for arbitrary $h(s)$, if
\begin{equation}
\label{QC_no_h}
\frac{2B'(L,s)}{1 - B(L,s)} + \frac{B''(L,s)}{B'(L,s)} \geq 0.
\end{equation}

By substituting $B'(L,s) = B^2(L,s) + \left( \frac{L}{s} -1 \right) B(L,s)$ and $B''(L,s) = 2B^3(L,s) + 3\left(\frac{L}{s}-1 \right) \times B^2(L,s) +\left[\left(\frac{L}{s}-1 \right)^2-\frac{L}{s^2} \right] B(L,s)$ into Eq. (\ref{QC_no_h}), we may rewrite Eq. (\ref{QC_no_h}) as
\begin{equation}
\label{QC_B}
d_L(s) \times \frac{B(L,s)}{[1-B(L,s)]B'(L,s)\left( \sum_{i=0}^{L}\frac{s^{i}}{i!}\right)^2s^2}  \geq 0,
\end{equation}
where $d_L(s)$ is a polynomial given by $d_L(s)=\sum_{n=0}^{2L-1} a_n s^n$. For $L=1,\ldots,5$, it is easy to check that $a_n \geq 0$ and therefore Eq. (\ref{QC_B}) holds for $s \geq 0$. For $L \geq 6$, the coefficient $a_n$ can be presented by
\begin{equation}
\label{QC_Poly}
a_n =
\begin{cases}
L^2-L,                                                   &  n=0,\\
2L^2-4L,                                                 &  n = 1,\\
\frac{2^n\left[ \frac{n^2+n}{4}+L^2-(n+1)L\right]}{n!},  &  2\le n \le L-1,\\
\frac{2^{L-2}(L-3)+L-1}{(L-1)!},                         &  n=L,\\
\sum_{ i=n-L+1}^{L-1}\frac{(n-L-i)(n-L-i-1)-2L}{i!(n-i)!} +\frac{4L-2n-2}{(L-1)!(n-L)!},                           &  L+1\le n\le 2L-5,\\
\frac{2}{(L-1)!(L-2)!}                                   &  n = 2L-4.
\end{cases}
\end{equation}

Again it is easy to check that $a_n \geq 0$ for $n \leq L$ and $n = 2L-4$. Note from Eq. (\ref{QC_D}) that $\Delta_L(l)=a_{L+l}$ for given $L$ and $l=1,2,\ldots,L-5$. If $\Delta_L(l) \geq 0$, Eq. (\ref{QC_B}) holds and hence $ h_L(s)$ is quasi-concave.
\end{IEEEproof}

Based on Theorem \ref{thm_QC_D}, we further present the following corollary.
\begin{corollary}
\label{cor_QC_100}
Problem (\ref{r_finite}) is a quasi-convex optimization if $L \leq 1000$.
\end{corollary}
\begin{IEEEproof}
We calculate $\Delta_L(l)$ for $L \leq 1000$, all of which are greater than or equal to zero. By noting that $u \leq 1$, we may conclude that $r(L,s)$ is quasi-concave.\footnote{Theoretically, we are capable of testing the quasi-concavity for any given $L$ using a computer. By observing the numerical results of $\Delta_L(l)$, we propose a conjuncture that $h_L(s)$ and $r(L,s)$ is quasi-concave for $L \geq 1$.}
\end{IEEEproof}

For quasi-convex optimization (\ref{r_finite}), binary search converges to the global optimum and thus can be adopted to further reduce the complexity of caching with finite buffer.


\begin{thebibliography}{1}

\bibitem{Leung}
X. Wang, M. Chen, T. Taleb, A. Ksentini, and V. Leung, "Cache in the air: exploiting content caching and delivery techniques for 5G systems," \emph{IEEE Communications Magazine}, vol. 52, no. 2, pp. 131-139, Feb. 2014.

\bibitem{Kato}
S. Arai, Z. M. Fadlullah, T. Ngo, H. Nishiyama, and N. Kato, "An efficient method for minimizing energy consumption of user equipment in storage-embedded heterogeneous networks," \emph{IEEE Wireless Communications}, vol. 21, no. 4, pp. 70-76, Aug. 2014.

\bibitem{Sherman}
S. Zhang, N. Zhang, X. Fang, P. Yang, and X. Shen, "Self-sustaining caching stations: Toward cost-effective 5G-enabled vehicular networks," \emph{IEEE Communications Magazine}, vol. 55, no. 11, pp. 202-208, Nov. 2017.

\bibitem{Social}
F. Malandrino, M. Kurant, A. Markopoulou, C. Westphal, and U. C. Kozat, "Minimizing peak load from information cascades: Social networks meet cellular networks,"  \emph{IEEE Transactions on Mobile Computing}, vol. 15, no. 4, pp. 895-908, April 2016.

\bibitem{Debbah_Bigdata}
E. Zeydan, E. Bastug, M. Bennis, M. A. Kader, I. A. Karatepe, A. S. Er, and M. Debbah, "Big data caching for networking: moving from cloud to edge," \emph{IEEE Communications Magazine}, vol. 54, no. 9, pp. 36-42, Sept. 2016.

\bibitem{Recommend}
L. E. Chatzieleftheriou, M. Karaliopoulos, and I. Koutsopoulos, "Caching-aware recommendations: Nudging user preferences towards better caching performance," \emph{Proc. IEEE International Conference on Computer Communications (INFOCOM)}, May 2017.

\bibitem{W.Chen_WCM}
Q. Yan, W. Chen, and H. V. Poor, "Big data driven wireless communications: A human-in-the-loop pushing technique for 5G systems," \emph{IEEE Wireless Communications}, vol. 25, no. 1, pp. 64-69, Feb. 2018.

\bibitem{3C_Synergy}
S. Andreev, O. Galinina, A. Pyattaev, J. Hosek, P. Masek, H. Yanikomeroglu, and Y. Koucheryavy, "Exploring synergy between communications, caching, and computing in 5G-grade deployments," \emph{IEEE Communications Magazine}, vol. 54, no. 8, pp. 60-69, Aug. 2016.

\bibitem{3C_Cornerstone}
E. K. Markakis, K. Karras, A. Sideris, G. Alexiou, and E. Pallis, "Computing, caching, and communication at the edge: The cornerstone for building a versatile 5G ecosystem," \emph{IEEE Communications Magazine}, vol. 55, no. 11, pp. 152-157, Nov. 2017.

\bibitem{M.Ali}
M. A. Maddah-Ali and U. Niesen, "Fundamental limits of caching," \emph{IEEE Transactions on Information Theory}, vol. 60, no. 5, pp. 2856-2867, May 2014.

\bibitem{Ali_Nonuniform}
U. Niesen and M. A. Maddah-Ali, "Coded caching with nonuniform demands," \emph{IEEE Transactions on Information Theory}, vol. 63, no. 2, pp. 1146-1158, Feb. 2017.

\bibitem{Diggavi}
J. Hachem, N. Karamchandani, and S. N. Diggavi, "Coded caching for multi-level popularity and access," \emph{IEEE Transactions on Information Theory}, vol. 63, no. 5, pp. 3108-3141, May 2017.

\bibitem{Approx_Tradeoff}
A. Sengupta and R. Tandon, "Improved approximation of storage-rate tradeoff for caching with multiple demands," \emph{IEEE Transactions on Communications}, vol. 65, no. 5, pp. 1940-1955, May 2017.

\bibitem{Gunduz_Improve}
M. M. Amiri and D. G\"und\"uz, "Fundamental limits of coded caching: Improved delivery rate-cache capacity tradeoff," \emph{IEEE Transactions on Communications}, vol. 65, no. 2, pp. 806-815, Feb. 2017.

\bibitem{W.Choi}
B. Hong and W. Choi, "Optimal storage allocation for wireless cloud caching systems with a limited sum storage capacity," \emph{IEEE Transactions on Wireless Communications}, vol. 15, no. 9, pp. 6010-6021, Sept. 2016.

\bibitem{B.Ottersten}
T. X. Vu, S. Chatzinotas, and B. Ottersten, "Coded caching and storage planning in heterogeneous networks," \emph{Proc. IEEE Wireless Communications and Networking Conference (WCNC)}, Mar. 2017.

\bibitem{Caire}
M. Ji, G. Caire, and A. F. Molisch, "Wireless device-to-device caching networks: Basic principles and system performance," \emph{IEEE Journal on Selected Areas in Communications}, vol. 34, no. 1, pp. 176-189, Jan. 2016.

\bibitem{G.Cao}
L. Qiu and G. Cao, "Popularity-aware caching increases the capacity of wireless networks," \emph{Proc. IEEE International Conference on Computer Communications (INFOCOM)}, May 2017.

\bibitem{H.Liu}
H. Feng, Z. Chen, and H. Liu, "Performance analysis of push-based converged networks with limited storage," \emph{IEEE Transactions on Wireless Communications}, vol. 15, no. 12, pp. 8154-8168, Dec. 2016.

\bibitem{A.Eryilmaz}
J. Tadrous and A. Eryilmaz, "On optimal proactive caching for mobile networks with demand uncertainties," \emph{IEEE/ACM Transactions on Networking}, vol. 24, no. 5, pp. 2715-2727, Oct. 2016.

\bibitem{W.Chen_TCOM1}
W. Chen and H. V. Poor, "Content pushing with request delay information," \emph{IEEE Transactions on Communications}, vol. 65, no. 3, pp. 1146-1161, March 2017.

\bibitem{W.Chen_TCOM2}
Y. Lu, W. Chen, and H. V. Poor, "Multicast pushing with content request delay information," \emph{IEEE Transactions on Communications}, vol. 66, no. 3, pp. 1078-1092, March 2018.

\bibitem{W.Chen_Globecom}
W. Chen and H. V. Poor, "Caching with statistical request delay information," \emph{Proc. IEEE Global Communications Conference (GLOBECOM)}, Dec. 2017.

\bibitem{Retention}
S. Shukla and A. A. Abouzeid, "Proactive retention aware caching," \emph{Proc. IEEE International Conference on Computer Communications (INFOCOM)}, May 2017.

\bibitem{Little}
J. D. C. Little, "A proof for the queuing formula: $L = \lambda W$," \emph{Operations Research}, vol. 9, no. 3, pp. 383-387, June 1961.

\bibitem{Zorich}
V. A. Zorich, \emph{Mathematical Analysis}, vol. I. 2nd Edition, Springer, 2015.

\bibitem{Whitt}
W. Whitt, "Heavy-traffic approximations for service systems with blocking," \emph{AT\&T Bell Laboratories Technical Journal}, vol. 63, no. 5, pp. 689-708, May-June, 1984.

\bibitem{Borovkov}
A. A. Borovkov, "On limit laws for service processes in multi-channel systems," \emph{Siberian Mathematical Journal}, pp. 746-763, 1967.

\bibitem{Convex}
S. Boyd and L. Vandenberghe, \emph{Convex Optimization}. Cambridge University Press, 2004.

\end{thebibliography}
\end{document}